\documentclass[reprint, floatfix, amssymb, amsfonts, nofootinbib, prb]{revtex4-1}

\usepackage[latin1]{inputenc}
\usepackage{dcolumn}
\pdfoutput=1
\usepackage{graphicx}
\usepackage{amsmath, amssymb, mathrsfs}
\usepackage{icomma}
\usepackage[caption=false]{subfig}
\usepackage{bm}
\usepackage{natbib}
\usepackage{epstopdf}
\usepackage{placeins}
\usepackage{hyperref}
\hypersetup{
    colorlinks=true,
    linkcolor=blue,
    citecolor=blue,
    urlcolor=blue,
    citebordercolor=1 1 1,
    linkbordercolor=1 1 1,
    menubordercolor=1 1 1,
    pagebordercolor=1 1 1,
    urlbordercolor=1 1 1,
  }

\DeclareMathOperator{\e}{e}

\begin{document}

\title{Monte Carlo Simulation of the Crossover from Bose Glass to Bragg Glass Phase in Layered BSCCO with Columnar Defects}
\author{L. M. Queiroz}
\author{M. D. Coutinho-Filho}
\affiliation{Laborat\'orio de F\'i­sica Te\'orica e Computacional$,$ Departamento de F\'i­sica$,$ Universidade Federal de Pernambuco$,$ 50670-901$,$ Recife-PE$,$ Brazil}

\begin{abstract}
 Monte Carlo simulations of layered BSCCO samples are used to investigate the behavior of vortex matter at low fields, particularly in connection with the possible occurrence of a Bragg glass (BrG) phase at low density of columnar defects, a phenomenon characterized by the prevalence of short-range over long-range order. In this dislocation-free topological phase the translational order correlation function displays a power law decay. For magnetic induction $B=0.1$ kG the analysis of the data for the first Bragg peak of the planar structure factor, the hexatic order parameter, and the Delaunay triangulation shows that, as the density of columnar defects is lowered, a \textit{crossover} (or transition) from Bose glass to BrG phase takes place in this \textit{highly anisotropic} high-T${}_c$ superconductor. Most importantly, an analysis of the {low-temperature} 3D vortex-vortex correlation function in terms of the structure factor, calculated via a saddle point approach and the use of the numerical data as input, provides clear-cut evidence of {the} power law decay of the {divergent} Bragg peaks in the BrG phase, a fundamental feature that was inequivocally verified only in isotropic compounds.
\end{abstract}

\maketitle

\section{Introduction\label{sec:1}}

Vortex matter in high-Tc superconductors has been a topic of intensive research \cite{blatter:1994, nelson:2002, ledoussal:2010, rosenstein:2010}. In these systems, the combined effect of magnetic interactions and Josephson coupling between vortices in neighboring planes, thermal agitation, magnetic induction, and disorder, gives rise to very rich phase diagrams. In fact, the understanding of the effects of point disorder and columnar defects has offered a great diversity of experimental and theoretical challenges, particularly those related to the occurrence and properties of glass phases, such as, the vortex glass (VG) \cite{fisher:1989a, fisher:1991}, the Bose glass (BG) \cite{nelson:2002, nelson:1992, *nelson:1993}, and Bragg glass (BrG) \cite{ledoussal:2010, nattermann:1990, giamarchi:1994, *giamarchi:1995} phases. Most importantly, disorder tends to pin vortices, thus preventing dissipation under an applied electrical current \cite{blatter:1994, nelson:2002, ledoussal:2010, rosenstein:2010, bardeen:1965}, a phenomenon common to type-II superconductors, particularly cuprates under columnar defects: YBCO \cite{civale:1991, *konczykowski:1991},  BSCCO \cite{gerhauser:1992, *thompson:1992}, and TBCCO \cite{budhani:1992}; iron-based \cite{kordyuk:2012, hosono:2015}, and multi-band superconductors in general \cite{drocco:2013, xu:2014}. Facing these challenges has required high-quality samples with disorder control, and the use of a variety of powerful experimental and theoretical techniques. Pursuing this endeavor, many discoveries and new concepts and mechanisms underlying the related intriguing phenomena have been continued uncovered and put forward. Recent advances include the study of single vortex unzipping (manipulation) using magnetic-force microscopy for extended and point defects \cite{kafri:2007, auslaender:2009}, the probe of vortex dynamics and pinning of single vortex in superconductors at nanometer scales \cite{embon:2015}, and relaxation dynamics of vortex lines in disordered type-II superconductors under magnetic field and temperature quenches \cite{assi:2015}. 

In this work, we focus on cuprate compounds, whose discovery opened the ``Pandora's box'' thirty years ago: LBCO \cite{bednorz:1986}, YBCO \cite{wu:1987}, BSCCO \cite{maeda:1988, *subramanian:1988}, and TBCCO \cite{sheng:1988a, *sheng:1988b}. In clean samples, we highlight the observation of the first-order melting transition of the Abrikosov lattice in BSCCO \cite{zeldov:1995} and YBCO \cite{schilling:1996, *schilling:1997}, and its description through the elastic approach \cite{brandt:1995}, the so-called boson analogy \cite{nelson:2002, nelson:1988, *nelson:1989, nordborg:1997} and Monte Carlo (MC) simulations \cite{blatter:1994, nelson:2002, ledoussal:2010, rosenstein:2010}, including the wandering of vortex lines in the solid phase (before melting), and the entanglement, cutting and reconnection of these vortices in the liquid phase (VL). Moreover, these compounds exhibit intermediate or high anisotropy and the phenomenon of decoupling of the CuO${}_2$ layers leading to pancake vortices (point vortices in the 2D CuO${}_2$ layers) at sufficiently high field (magnetic induction) \cite{schilling:1993, koshelev:1996, *koshelev:1997}. Despite challenging aspects of the BSCCO $B-T$ phase diagram \cite{fuchs:1998a, *shibauchi:1999, ando:1999, torres:2003}, the occurrence of vortex matter at intermediate ($0.1\text{ T} < B <1$ T) and high fields ($1\text{ T} < B < 10$ T), and the decoupling of the CuO${}_2$ planes in high fields at coincidence with the first-order melting transition \cite{torres:2003}, were verified also through {MC} simulations \cite{viana:2004} using the Lawrence-Doniach (LD) model \cite{ryu:1992, *hellerqvist:1994}, {the latter} derived from the anisotropic Landau-Ginzburg model with the neighboring 2D CuO${}_2$ layers coupled via the Josephson interaction. 

At this point, we find it is instructive to digress on some fundamental aspects of the BrG, VG, and BG phases, including the contrast and similarities caused by the presence of point disorder or columnar defects. It will prove useful, since our main goal is to study the crossover (or transition) from the BG to the BrG phase, under columnar defects, and to provide clear-cut evidence of diverging Bragg peaks in the BrG phase in highly anisotropic superconductors, such us BSCCO, under low fields and low density of columnar defects. Indeed, this issue remains challenging \cite{ledoussal:2010}, since this fundamental feature was suggested \cite{giamarchi:1995} to explain neutron diffraction data in 2H-NbSe${}_2$ \cite{yaron:1994}, an isotropic compound, and, most unequivocally, verified  \cite{klein:2001} only on the isotropic single-phase (K, Ba) BiO${}_3$- crystal, under low density of point disorder.

\textit{Point Disorder} (BrG and VG): In the context of elastic theory the effect of point disorder on the Abrikosov lattice can be classified in three main regimes \cite{ledoussal:2010}. The Larkin-Ovchnnikov regime \cite{larkin:1970, *larkin:1979}, in which case the linear coupling of the displacement field to the disorder causes the breaking of the system into independently pinned domains. The random manifold regime sets in at larger scales \cite{feigelman:1989}, with the decay of the translational order correlation function governed by a stretched exponential \cite{feigelman:1989, bouchaud:1991, *bouchaud:1992}, instead of the pure exponential decay of the Larkin regime, and verified in dislocation free samples of BSCCO \cite{kim:1999}. Lastly, in the asymptotic regime, displacements grow logarithmic at large scales  \cite{nattermann:1990, giamarchi:1994, *giamarchi:1995} and the translational order correlation function displays a power law decay. This regime is the so-called BrG phase, a dislocation-free topological phase. Dynamic features of this phase were confirmed in BSCCO \cite{fuchs:1998b}, including the moving BrG behavior \cite{giamarchi:1996, *giamarchi:1998}, both experimentally \cite{pardo:1998} and numerically \cite{chen:2003}.

In addition, the increase of point disorder or the increase of the magnetic induction $B$ can drive a first-order transition from the low-temperature BrG phase to the VG phase \cite{fisher:1989a, fisher:1991, lidmar:2003}, or to a vortex liquid (VL) phase at higher temperatures, as numerically and experimentally verified \cite{blatter:1994, ledoussal:2010, rosenstein:2010}. Indeed, a detailed experimental study of the vortex state in La${}_{1.9}$Sr${}_{0.1}$CuO${}_4$ at a macroscopic level reported a change with field from BrG to VG \cite{divakar:2004}, in which case the microscopic behavior reflects a delicate interplay of thermally induced and pinning-induced disorder. A VG will exhibit off-diagonal long-range order, much in analogy with a spin glass phase, and a single pinned vortex is formally equivalent to a direct polymer in a random media \cite{fisher:1991}. In fact, as disorder or $B$ increases, dislocations start to proliferate, and the occurrence of a VG phase is experimentally verified by the abrupt change of the Josephson plasma frequency by crossing the BrG-VG first-order transition line (the BrG-VL Transition is also first order) \cite{gaifullin:2000}. Notwithstanding, experimental groups have struggled hardly in order to obtain clear-cut $B-T$ phase diagrams \cite{beidenkopf:2005, *beidenkopf:2007} in light of the VG and VL phases mentioned above. In particular, an unusual glassy state (pinned vortex liquid) at intermediate fields was recently reported  \cite{heron:2013}; in fact, this glass state freezes continuously from the equilibrium VL, but its structure differs from both the low-field BrG and the high-field VG phases.

\textit{Columnar defects} (BG and BrG): Let us now focus on the effect of columnar or correlated disorder, experimentally motivated by the sizable upward shift of the irreversibility line in heavy-ion irradiated samples of YBCO \cite{civale:1991, konczykowski:1991}, BSCCO \cite{gerhauser:1992, *thompson:1992},and TBCCO \cite{budhani:1992}. Indeed, an extension of the boson analogy \cite{nelson:1988, *nelson:1989} allowed the mapping \cite{nelson:2002, nelson:1992, *nelson:1993} of line vortex under columnar defects onto the problem of two-dimensional disordered driven boson localization \cite{fisher:1989b}, thereby giving rise to the solid BG phase, the (superfluid) entangled flux liquid phase, and the Mott insulator phase, with one fluxon localized on every pin. It was emphasized \cite{nelson:1992, *nelson:1993} that, despite the similarities of the scaling laws for the BG and VG phases, point disorder promotes wandering and entanglement, while columnar disorder inhibits wandering and promotes localization. Direct contact with experimental observations in BSCCO under columnar defects \cite{dai:1994} was nicely provided \cite{tauber:1995a, *nelson:1996, tauber:1997, *tauber:1998}, including dynamical effects \cite{nelson:2002}. In addition, several pertinent aspects of the distinction between the regimes $B < B_\phi$, where the BG theory fully applies, and the more complex behavior expected for $B > B_\phi$, were reported \cite{radzihovsky:1995, tauber:1995, *wengel:1998, larkin:1995, nandgaokar:2002}; here, the matching field is defined by 
\begin{equation}
 B_\phi = n_d\Phi_0,
\end{equation}
where $n_d$ is the disorder concentration and $\Phi_0$ is the superconducting flux quantum. In particular, in contrast to the strong BG regime ($B < B_\phi$), in the weakly pinned BG regime ($B > B_\phi$) the occurrence of an interstitial liquid phase was proposed \cite{radzihovsky:1995} to occur between the former phase and the VL phase due to the increase of flux lines and thermal excitations, with the two BG regimes separated by the Mott insulator line at $B = B_\phi$ in a $B-T$ phase diagram. Detailed studies on samples of BSCCO and YBCO with columnar defects were undertaken \cite{vanderbeek:1995b, vanderbeek:2001, soret:2000} in order to clarify the diversity of estimates for the pertinent critical exponents\cite{lidmar:1999}, which contrast with those of the VG transition \cite{lidmar:2003}. In fact, for $B < B_\phi$ the data is well described by the BG theory for vortex pinning and dynamics in the presence of columnar defects, while in the regime $B > B_\phi$ the dynamics is most likely determined by the collective activation of pancake vortices from interstitial vortices not trapped by columnar defects.

Recently \cite{queiroz:2015}, {the} authors and a collaborator used the LD model to perform MC simulations of 3D layered samples with BSCCO parameters under columnar defects. The numerical data of the temperature behavior of the structure factor and vortex-vortex correlation length along the field ($z$-direction) brought clear-cut evidence of hysteretic behavior of the vortex matter at intermediate ($0.1\text{T} \leq B < 1$ T) and high fields ($1\text{ T} \leq B \leq 10$ T), and for $B_\phi/B = 1/4$, 1, and 4. For this purpose, two representative initial conditions at zero temperature ($T$) were used: the Abrikosov lattice and a random vortex lattice, mimicking possible configurations in a zero-field-cooled (ZFC) protocol. At intermediate fields and increasing temperature, we observed that both ZFC configurations evolve through metastable states and meet the pristine melting transition line $B_m(T)$, thus defining the melting temperature $T_m$ for a given B. We also verified that, in both cases, the vortex matter undergoes a smooth plane decoupling transition (formation of pancake-like vortex structure) around $T_m$, with the correlation length along the z-direction characterized by a T-dependent exponential decay with z. In this regime, we can also visualize vortex pinning to one or two neighboring columnar defects at the onset of vortex depinning, and entanglement between two vortices above melting. In addition, the very relevant case of the field-cooling (FC) process from an initial temperature $T_0 = 79$ K at the VL phase is also considered. In this case, as T decreases under intermediate field values, the system evolves through metastable states of an inhomogeneous phase of unpinned vortices coexisting with pinned ones  (BG background). Lastly, the system reaches a FC robust BG phase down to very low T. We stress that, under the above-mentioned conditions, the melting and decoupling scenario for the ZFC and FC protocols are practically identical. On the other hand, for high fields and under the ZFC protocols, the melting transition is practically concurrent with the discontinuous decoupling of the CuO${}_2$ planes; while, under FC, the vortex lattice decouples at a temperature below the melting transition. Indeed, under FC, we identify that the exchange between flux lines is the underlying mechanism for plane decoupling and the formation of a pancake-like vortex structure.

At this stage, it is worth mentioning that the corresponding melting and BG lines at intermediate and high fields mentioned above \cite{queiroz:2015} can be incorporated quite successfully in the B-T phase diagram of BSCCO samples \cite{fuchs:1998a, *shibauchi:1999, torres:2003}. They are also compatible with phase diagrams (pin concentration versus temperature) derived using numerical functional minimization techniques of electromagnetic interactions between vortices \cite{dasgupta:2003, *dasgupta:2004, dasgupta:2005, dasgupta:2009} and MC simulations of the 3D frustrated anisotropic XY model \cite{nonomura:2004}  (see phase diagrams in Figs. 1 of both references \onlinecite{dasgupta:2005} and \onlinecite{nonomura:2004}), apart from details inherent to the use of distinct approaches and choice of parameters. Moreover, MC simulations using the LD model for BSCCO at $B = 125$ G, and $B_\phi/B = 1/5$,\cite{tyagi:2003} found that the vortex matter displays an increase of the magnitude of the structure factor at the first Bragg peak and of the line wandering along the $z$ direction, just before the transition to the IL phase, consistent with our results at intermediate fields \cite{queiroz:2015}.

We also remark that the melting scenario at low-fields observed in BSCCO for $B\leq200$ G and $5\text{ G}\leq B_\phi \leq100$ G \cite{banerjee:2003, *menghini:2003, *banerjee:2004}, including dynamical effects and oblique fields \cite{avraham:2007}, have also identified an intermediated phase which resembles the one found in Ref. \onlinecite{queiroz:2015} at intermediate and high fields. Indeed, the low-field scenario evidences that the delocalization (or depinning) line, separates the homogeneous VL phase from an inhomogeneous one in which nanodroplets of vortex liquid are caged in the pores of a solid skeleton formed by vortices pinned on columnar defects (porous vortex matter). It is also verified that all pertinent lines merge to the low-field pristine melting line $B_m^0(T)$, which ends at $B_m^0(T_c)=0$ This scenario is also consistent with analytical \cite{lopatin:2004, *kierfeld:2005} and appropriate numerical modeling \cite{goldschmidt:2005, *goldschmidt:2007}. The above features are very important in the context of identifying the nature of phases in the $B-T$ phase diagram of BSCCO, as discussed for clean samples \cite{viana:2004} in the second paragraph of this section, and extended in Ref. \onlinecite{queiroz:2015}, and in this manuscript, to include the effect of columnar disorder.

In the next sections we shall {present} our numerical and analytic studies with focus on the main goals of our work, namely, the numerical-analytic description of the crossover (or transition) from the BG to the BrG phase, under columnar defects, and the clear-cut evidence of diverging Bragg peaks in the BrG phase in a highly anisotropic superconductor with BSCCO parameters, under low fields and low density of columnar defects. In Sec. \ref{sec:2} we describe the LD model, suitable to describe highly anisotropic BSCCO samples, the phenomenological physical quantities, and the simulation procedure. In Sec. \ref{sec:3} we present the results related to both hexatic order parameter and the first Bragg peak of the structure factor for the vortex matter at different disorder concentrations, supplemented by the Delaunay triangulation of the vortices. Quite rewarding, we find that, at $B$ = 100 G and $B = B/32$, the in-plane structure factor displays sharp Bragg peaks hexagonally distributed on the lattice; in fact, we observe that the system undergoes a crossover from a BG to a BrG phase as $B$ is lowered from $B_\phi = B$ to $B_\phi =B/32$. In Sec. \ref{sec:4} we examine the 3D vortex-vortex correlation function in terms of its Fourier transform, i. e., the 3D structure factor. A saddle point calculation, and the use of the numerical data as input, clearly demonstrates the occurrence of divergent Bragg peaks for simulated BSCCO samples in a BrG phase under low fields and low density of columnar defects. Lastly, Sec. \ref{sec:5} is devoted to our concluding remarks.

\section{Modeling BSCCO with columnar defects\label{sec:2}}

We consider that the flux lines in a layered superconductor is an array of point-like pancake vortices bound together by an interlayer Josephson interaction \cite{ryu:1992, *hellerqvist:1994,viana:2004, viana:2006}. 

{The} following parameters, adequate for BSCCO {are used}\cite{blatter:1994}: $d=15\text{\AA}$, $s=1.66\text{\AA}$, $\lambda_0=1414.2\text{\AA}$, $T_c=87 K$, $\xi_{ab}(0)=21\text{\AA}$, $\gamma=100$, where $d$ is the interlayer spacing, $s$ is the thickness of a layer, $\lambda_0$ is the penetration depth at $T=0$, such that $\lambda_{ab}(T)=\lambda_0(1-T/T_c)^{-1/2}$, $\xi_{ab}(0)$, is the coherence length in the $ab$ plane at $T=0$, where $T_c$ is the the zero-field critical temperature, and $\gamma=\lambda_{z}/\lambda_{ab}=\sqrt{2}\xi_{ab}(0)/(\sqrt{g}d)$ measures the ratio between axial and planar penetration depths, where $g$ is the interlayer Josephson coupling strength. {Lastly, we know that: the average vortex-vortex distance is \cite{tinkham:2004} $a_0=\sqrt{2/\sqrt{3}}\sqrt{\Phi_0/B}$, where $\Phi_0=hc/2e\approx2.068\times10^{-15}\text{Tesla}\cdot \text{m}^2$. Thereby, in the low temperature regime, i. e., $T/T_c\ll1$, $a_0> \lambda_{ab}$ (see $\lambda_{ab}(T)$ above); while, as $T\rightarrow T_c$, the latter inequality can be reversed.}

We use the Lawrence-Doniach model, in which the system is represented by a stack of superconducting planes. Each plane has a certain number of interacting vortices, whose free energy is given by \cite{ryu:1992, *hellerqvist:1994, viana:2004, viana:2006}
\begin{align}\label{eq:2}
F_{LD}=& \frac{1}{8\pi}\int d^3r \mathbf{B}^2(r)+\frac{dH_c^2}{8\pi}\sum_{z=1}^{L_z}\int d^2\rho\left\{\left(1-\frac{T}{T_c}\right)\left|\psi_z\right|^2\right.\nonumber \\
&\left. +\frac{1}{2}\beta\left|\psi_z\right|^4+\left|\xi_{ab}(T)(\mathbf{\nabla}_{ab}-2ie\mathbf{A}_{ab})\psi_z\right|^2\right.\nonumber \\
&\left.-g\left|\exp\left(2ie\int_{z+d}^{z}dz\; A_z\right)\psi_{z+d}-\psi_{z}\right|^2\right\},
\end{align}
where $\mathbf{B} (r)$ is the local magnetic induction, $\psi_z$ denotes the dimensionless superconducting order parameter,  $H_c(T)$ is the thermodynamic critical magnetic field, $e$ is the electron charge, $\beta$ is the Landau coefficient of the quartic term in $\left|\psi_z\right|$, $\mathbf{\nabla}_{ab}$ is the in-plane gradient, $\mathbf{A}_{ab}$ ($A_z$) is the vector potential in the $ab$ plane (at the layer $z$). $\bm \rho$ is a position in the $ab$ plane. The above Helmholtz free energy is considered instead of the Gibbs one, since in our simulations we consider fixed $B$ (constant number of vortices) along the $z$ axis perpendicular to the superconducting $ab$ planes. It is worth mentioning that the conditions for a smooth connection \cite{tinkham:2004} between the LD model and the anisotropic Ginzburg-Landau model was properly discussed in Ref. \onlinecite{chapman:1995}, where rigorous aspects of the procedure were put forward.

Since our study is done in terms of a vortex representation, one can minimize Eq. (\ref{eq:2}) to obtain the various contributions to the energy of the system. Taking the London limit ($|\psi_z|$ constant throughout the sample), the Josephson interaction between a pair of adjacent layers can be expressed as \cite{bulaevskii:1992,ryu:1992,blatter:1994}
\begin{equation}\label{eq:3}
V_J= \frac{\Phi_0^2}{16\pi^3\lambda_{ab}^2\gamma^2d}\int d^2\rho\left[1-\cos\left(\tilde{\varphi}_{z+d}-\tilde{\varphi}_z\right)\right],
\end{equation}
where $\tilde{\varphi}_z$ is the gauge-invariant phase of $\psi_z$. The difference $\tilde{\varphi}_{z+d}-\tilde{\varphi}_z$ can be calculated for the case of one single pancake vortex dislocated from the flux line by a minimization procedure of Eq. (\ref{eq:2}). This leads to
\begin{equation}\label{eq:4}
  \nabla_{ab}^2(\tilde{\varphi}_{z+d}-\tilde{\varphi}_z)=\left(\frac{1}{\gamma^2d^2}+ \frac{1}{\lambda_z^2}\right)\sin(\tilde{\varphi}_{z+d}-\tilde{\varphi}_z).
\end{equation}
Considering $\lambda_z\gg\gamma d$, one can solve Eq. (\ref{eq:4}) and calculate the Josephson energy $V_J$. We define $\bm \rho_i(z)$ as the 2D vector position of the pancake vortex belonging to the $i$-th flux line at plane $z$. For vortices whose planar distance $r_1=|\bm \rho_i(z+d)-\bm \rho_i(z)|$ is much less than the length scale $\gamma d$, $V_J\sim (r_1/\gamma d)^2\ln(\gamma d/r_1)$. For $r_1\gg \gamma d$, the dependence of $V_J$ on $r_1$ is linear. Therefore, in our calculations, we have used the following expression for $V_J$:
\begin{align}\label{eq:5}
&\frac{d\Phi_0^2}{8\pi^3\lambda_{ab}^2}\left[1 + \ln \left(\frac{\lambda_{ab}}{d}\right)\right]\frac{r_1^2}{(\gamma d)^2}\ln \left(\frac{\gamma d}{r_1}\right), \; \text{if}\; r_1\leq \gamma d;\notag \\
&\frac{d\Phi_0^2}{8\pi^3\lambda_{ab}^2}\left[1 + \ln \left(\frac{\lambda_{ab}}{d}\right)\right]\left(\frac{r_1}{(\gamma d)} - 1\right),\; \text{otherwise}.
\end{align}
For vortices on the same plane, the magnetic repulsion can be given by the 2D Ginzburg-Landau model \cite{ryu:1992,blatter:1994}:
\begin{equation}\label{eq:6}
V_{plane}=\frac{\Phi_0^2s}{8\pi^2 \lambda_{ab}^2}K_0 \left(\frac{\vert {\bm \rho}_{ij}(z)\vert}{\lambda_{ab}}\right),
\end{equation}
where $\bm \rho_{ij}=|\bm \rho_i(z)-\bm \rho_j(z)|$.

{We  also introduce the random potential $V_D(\bm \rho_i(z))$, which emerges from $n_d$ random columnar defects located at a $z$-independent coordinate $\bm R_k$ via an attractive contact interaction (pinning) of constant magnitude $E_p=100$ K, with the vortex lines:}

\begin{equation}\label{eq:7}
 V_D(\bm \rho_i(z))=-E_p\sum_{k=1}^{n_d}\delta_{\bm \rho_i(z), \bm R_k},
\end{equation}
{where $\delta_{\bm \rho_i(z), \bm R_k}$ is the Kronecker delta.}

The energy to be minimized following the MC procedure discussed above reads:
\begin{align}\label{eq:8}
 E=&\sum_{i, z}V_D(\bm \rho_i(z))+\sum_{i, z}V_J(|\bm \rho_i(z+d)-\bm \rho_i(z)|)\notag \\
 &+\frac{1}{2}\sum_{i, j, z} V_{\text{plane}}(|\bm \rho_i(z)-\bm \rho_j(z)|).
\end{align}

Our approach allows us to calculate the first Bragg peak of the in-plane structure factor, $S(k_\text{Bragg})$, where $S(\bm k_\bot)=S(\bm k_\bot, z=0)$ is the planar structure factor in the momentum space, $\bm k_\bot=(k_x, k_y)$, {with}:
\begin{equation}\label{eq:9}
S({\bm k}_\bot, z)=\left[\frac{1}{\Omega N}\sum_{\rho} \langle n(0, 0)n({\bm \rho}, z)\rangle e^{i{\bm k}_\bot\cdot {\bm \rho}}\right]_{av},
\end{equation}
where $\langle n(0,0)n({\bm \rho}, z)\rangle$ is the {density-density} correlation function in plane $z$, $N$ is the number of lines, $\Omega$ is a normalization constant such that, for an Abrikosov lattice, $S(k_\text{Bragg})=1$, $n(\bm \rho, z)$ is the density of vortices at position $(\bm\rho, z)$, $\langle...\rangle$ denotes the thermal average, whereas $[...]_{av}$ stands for the process of sampling disorder average. Most importantly, we have also calculate the hexatic order parameter,
\begin{equation}\label{eq:10}
\Psi_6=\sum_{i=1}^N \frac{1}{z_i}\sum_{j=1}^6 \e^{6\textrm{i}\theta_{ij}},
\end{equation} 
where $\theta$ is the bond angle between next-neighbor vortices, and $z_i$ is the number of next-neighbors in the $i$th plane. Note that, for the hexagonal lattice, $\Psi_6=1$. Other pertinent correlation functions {have also been calculated, particularly} the root mean square deviation of vortices along the $z$ direction, {and} the 3D vortex-vortex correlation function and structure factor {(see Section IV)}.

{In Section III and IV, we report on results of Metropolis MC simulations on a system with 64 vortex lines on a grid of 256 $\times$ 222 $\times$ $N_z$, with $N_z=64$ layers and periodic boundary conditions in all directions. In our simulations each Monte Carlo step consists of one random change of each vortex position to one of the
neighboring sites on the lattice grid. The energy [Eq. \eqref{eq:8}] change is then calculated: if the change lowers the energy, we accept the move; else, there is a probability of accepting the move, or not, with the sum of these probabilities being equal to one according to the Metropolis algorithm adapted to Monte Carlo simulation of vortex matter. Further, we have checked that $2\times10^4$ steps are enough to equilibrate the data, and additional $10^5$ Monte Carlo Steps are taken for each temperature, with $\Delta T=\pm1$.}

\section{Hysteretic behavior and crossover from BG to BrG phase\label{sec:3}}

\begin{figure*}[htp]
\begin{center}
\begin{minipage}{0.32\textwidth}
  \subfloat{\includegraphics[width=\textwidth]{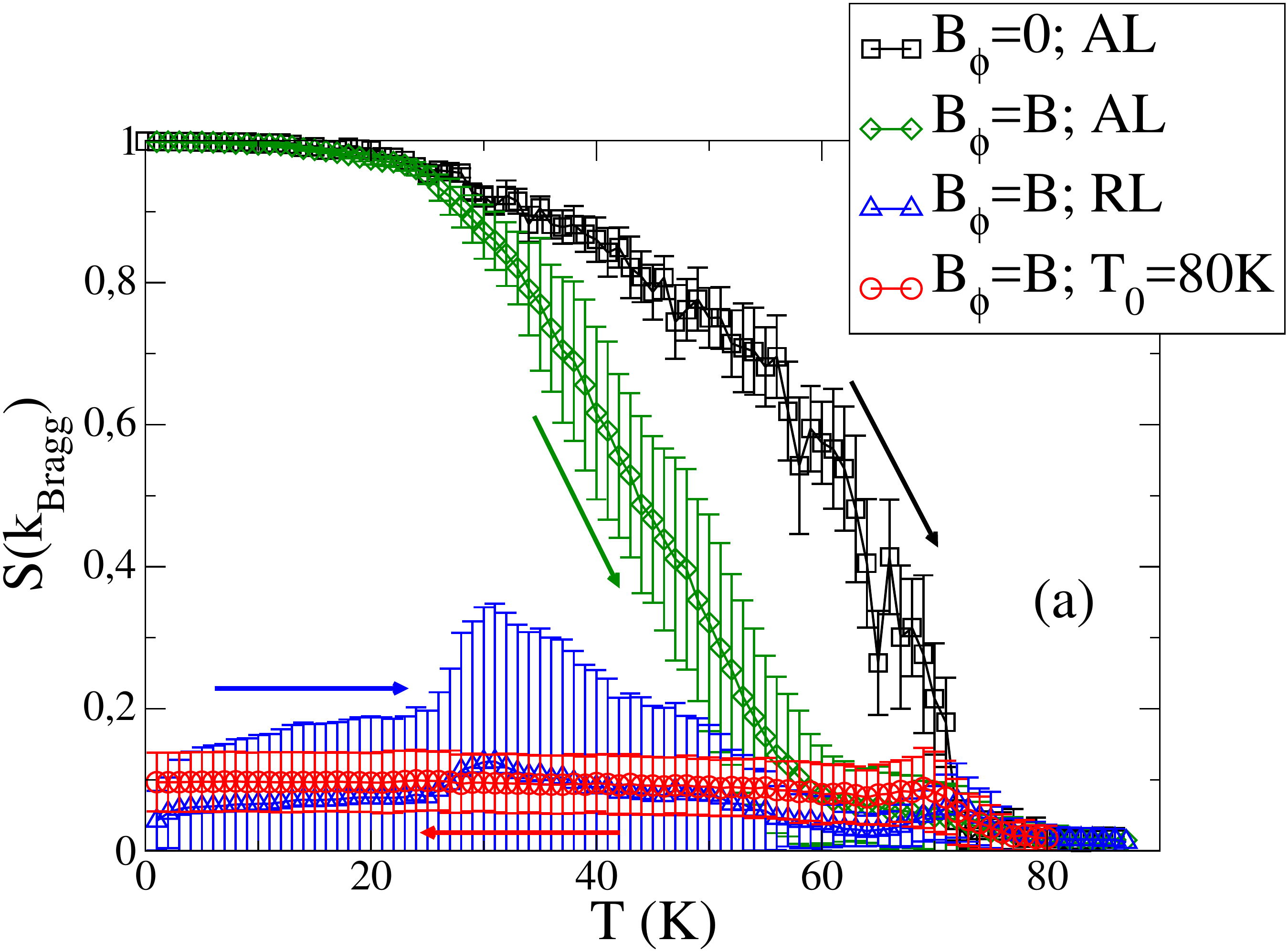}\label{fig1a}}\\
 \subfloat{\includegraphics[width=\textwidth]{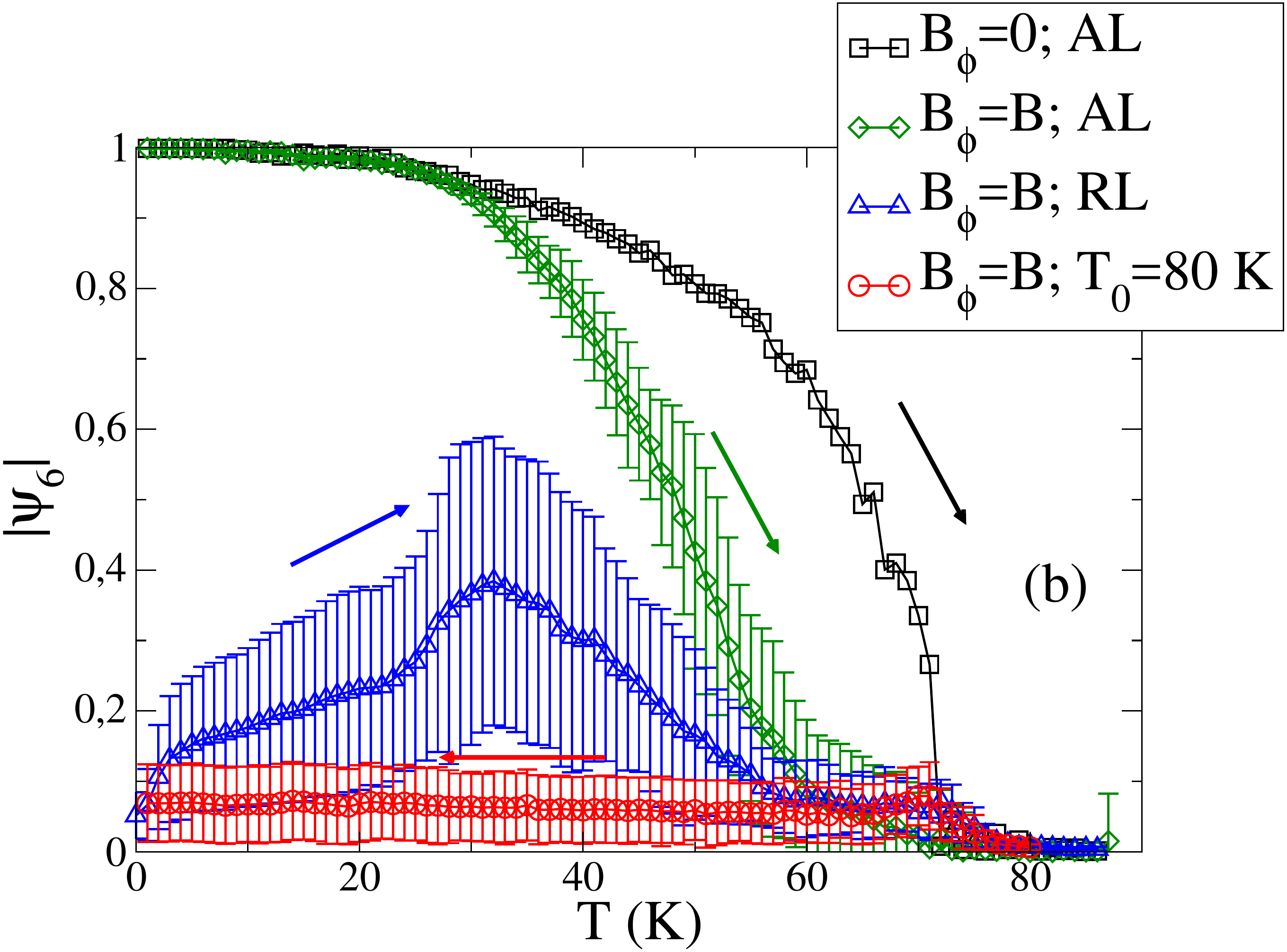}\label{fig1b}}
\end{minipage}
\begin{minipage}{0.32\textwidth}
 \subfloat{\includegraphics[width=\textwidth]{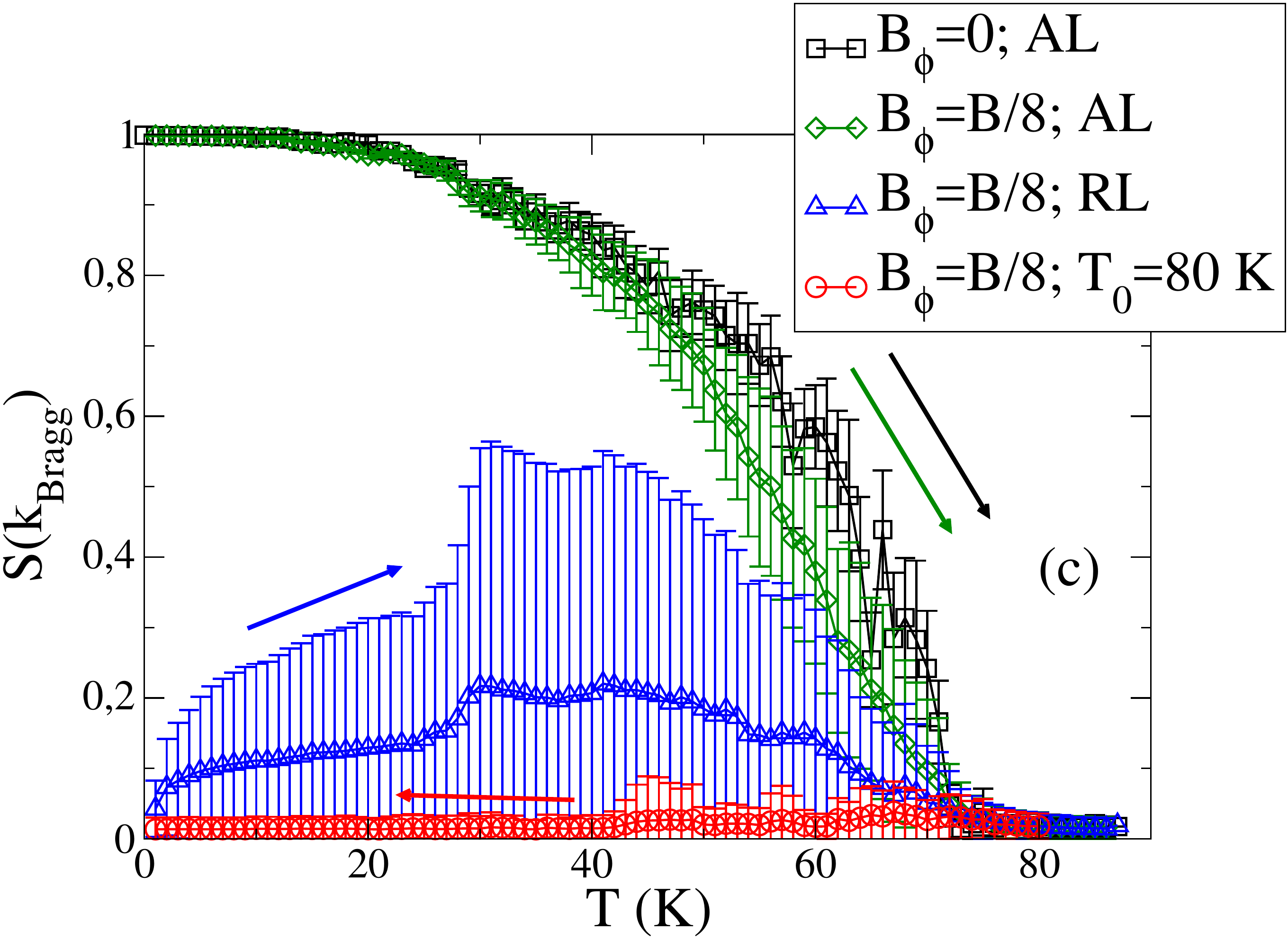}\label{fig1c}}\\
\subfloat{\includegraphics[width=\textwidth]{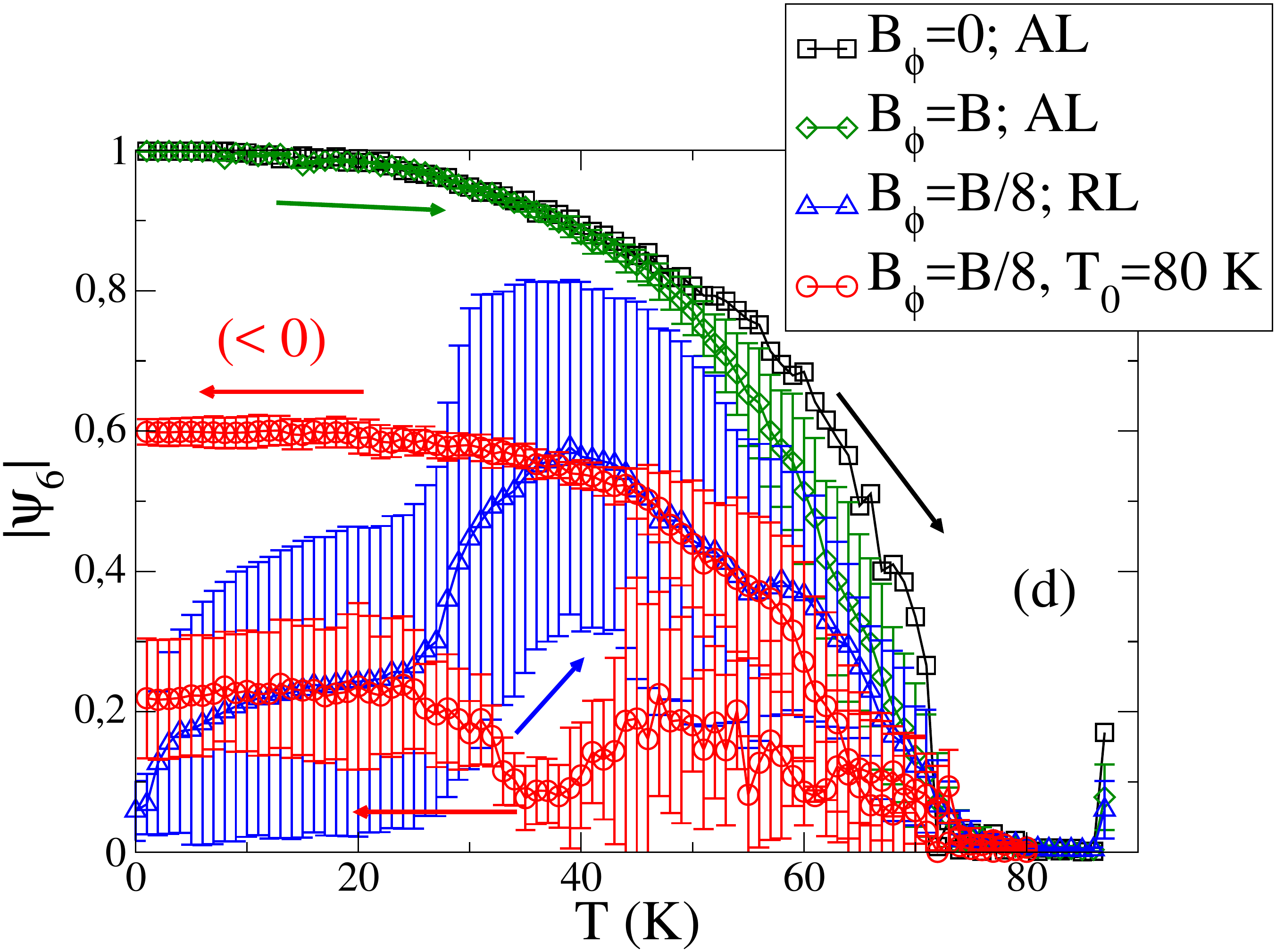}\label{fig1d}}
\end{minipage}
\begin{minipage}{0.32\textwidth}
\subfloat{\includegraphics[width=\textwidth]{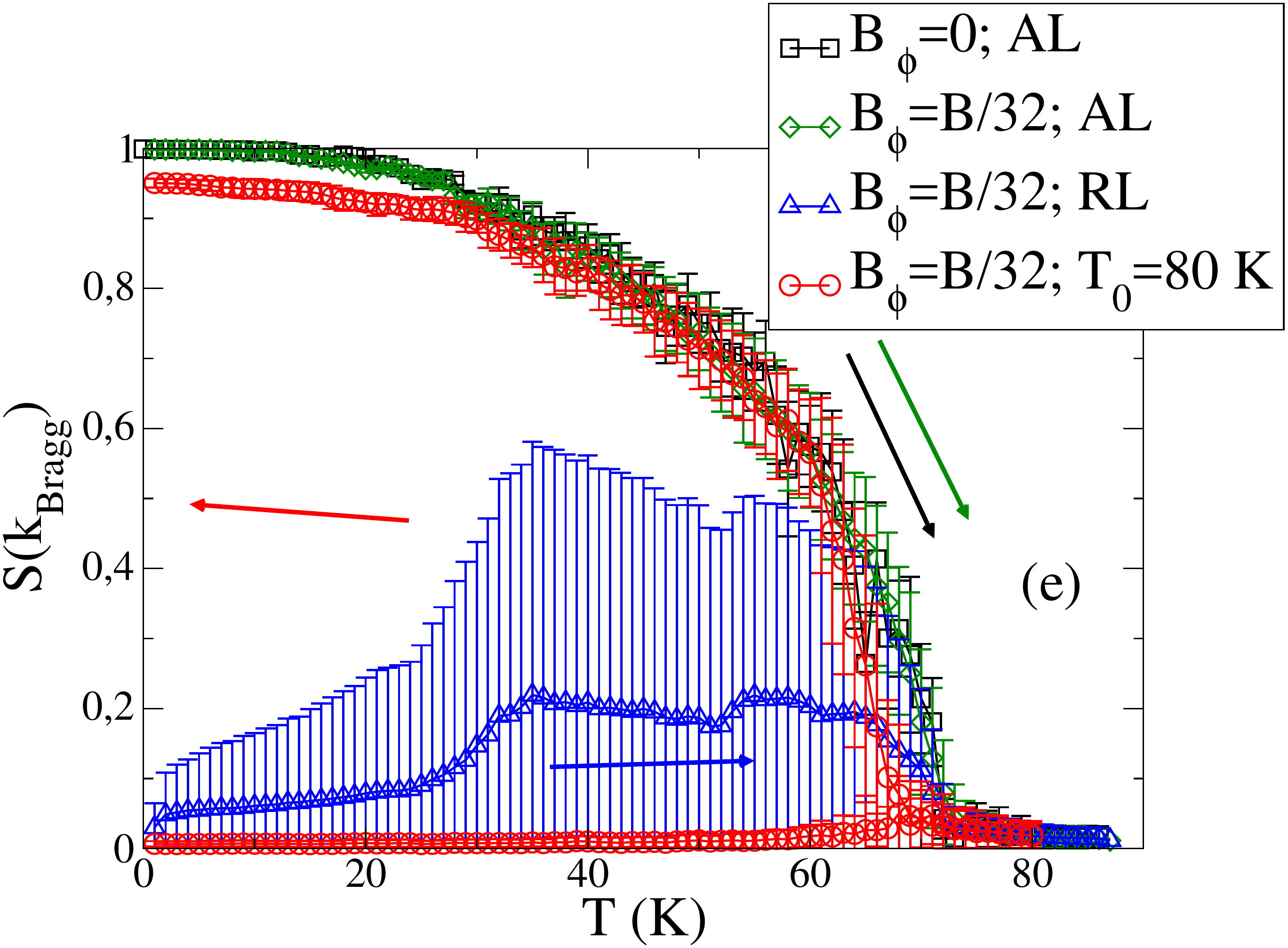}\label{fig1e}}\\
\subfloat{\includegraphics[width=\textwidth]{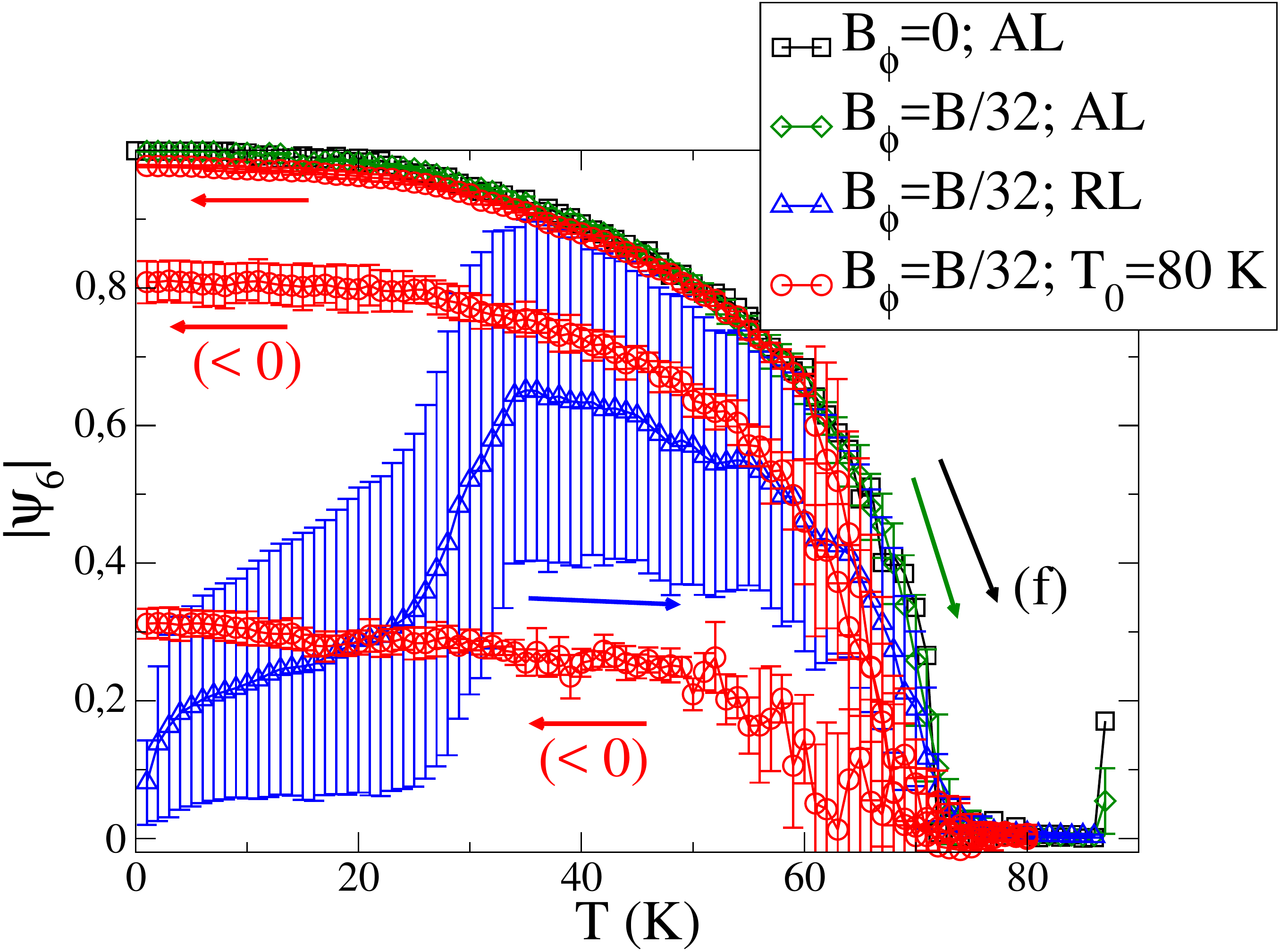}\label{fig1f}}
\end{minipage}
\caption{(Color online) First Bragg peak of the structure factor $S(k_\text{Bragg})$ [(a), (c) and (e)] and hexatic order parameter [(b), (d) and (f)], as a function of temperature for $B=0.1$ kG and $B_\phi=B$ [(a) and (b)], $B_\phi=B/8$ [(c) and (d)] and $B_\phi=B/32$ [(e) and (f)]. Data {of} pristine samples ($B_\phi=0$) are shown for comparison, including sampling fluctuations. For $B_\phi\neq0$ and distinct boundary conditions: Abrikosov lattice (AL), random lattice (RL), and field cooling with $T_0=80$ K; the sampling disorder average are indicated by a darker color curve (green, blue, and red). The arrows indicate data for increasing or decreasing temperature{, and the symbol ``$(<0)$'' in Figs. \protect\subref{fig1d} and \protect\subref{fig1f} indicates samples whose data of $\Psi_6$ exhibit negative values only.}\label{fig1}}
\end{center}
\end{figure*}

We have measured the structure factor and the hexatic order parameter of samples under $B=0.1$ kG and $B_\phi=B$, $B_\phi=B/8$, and $B_\phi=B/32$. We averaged these quantities over 30 samples for three initial conditions at $T=T_0$: Abrikosov and random lattices, both at $T_0=0$ with $T$ increasing {up to $T_\text{final}=87$ K} above melting; and a field-cooling (FC) protocol with $T_0=80$ K (above melting) with $T$ lowering down to $T_\text{final}=0$. 

Figure \ref{fig1} displays the first Bragg peak of the structure factor $S(k_\text{Bragg})$ [Figs. \subref{fig1a}, \subref{fig1c}, and \subref{fig1e}] and the hexatic order parameter $\Psi_6$ [Figs. \subref{fig1b}, \subref{fig1d}, and \subref{fig1f}] {of vortex matter under disorder, temperature and applied field conditions}.{The data present deviations from the average values and are a true manifestation of the complex behavior of the vortex matter under the referred conditions. In fact, larger deviations from the average are found only for the samples with random initial vortex configuration. This is so because of the ``frustrated'' tendency of the vortices to reach an Abrikosov lattice under this very awkward initial condition, as observed, for example, for $S(k_\text{Bragg})$ in Figs. 1(a), 1(c), and 1(e). In addition, through a careful examination of all 30 samples, we have identified that $\Psi_6$ in Figs. 1(b), 1(d) and 1(f) can be positive, negative, or to display both positive and negative values during the heating process up to the melting transition. Therefore, the average values for all data exhibited in Fig. \ref{fig1} were obtained using magnitude values of the pertinent quantity, thus allowing us to obtain smaller deviations from the average for several calculated quantities.}

Figures \subref{fig1a} and \subref{fig1b} exhibit $S(k_{\text{Bragg}})$ and $\Psi_6$, respectively, at the doping $B_\phi=B$. In Fig. \subref{fig1a}, for both Abrikosov and random initial lattice configurations, we can notice a distinct behavior in comparison with the data for strong and intermediate fields \cite{queiroz:2015}. Indeed, in the present case $S(k_{\text{Bragg}})$ intersects the curve for the pristine sample only at the melting transition (within {deviations from the average}){, while for $B=1$ kG \cite{queiroz:2015} the curves of $S(k_\text{Bragg})$ meet that of the pristine sample at $T\lesssim60$ K. Moreover, for $T\lesssim20$ K the data of the Abrikosov initial configuration do coincide with those of the pristine sample, which is not the case for the sample at $B=1$ kG in Fig. 1(b) of Ref. \onlinecite{queiroz:2015} that departs from the pristine sample at $T\gtrsim5$ K.}. As for the FC protocol, we see that, after cooling below the melting temperature, the system settles itself in a glassy configuration and remains there down to $T=0$. We identify this vortex ``freezing'' with the BG phase also observed in higher fields \cite{queiroz:2015}, although here we find a {much} smaller magnitude for the first Bragg peak, i. e.,  $S(k_\text{Bragg})\simeq 0.1${, instead of $S(k_\text{Bragg})\simeq 0.4$ found at the intermediate field value $B=1$ kG\cite{queiroz:2015}}. {Now, comparing Figs. \subref{fig1a} and \subref{fig1b}, we note that the average values of $S(k_\text{Bragg})$ and $\Psi_6$ display similar behavior, regardless of the initial condition, although for the FC protocol, $S(k_\text{Bragg})$ is more robust than $\Psi_6$}. {The above results suggest that this sample at $B=B_\phi=0.1$ kG is very close to the onset} of the {BG-BrG} crossover.

Figures \subref{fig1c} and \subref{fig1d} display $S(k_{\text{Bragg}})$ and $\Psi_6$ {at} $B_\phi=B/8$, respectively, for the same families of initial configurations analyzed in the case $B_\phi=B$.  Figure \subref{fig1c} shows that, for the Abrikosov initial configuration, {the average value of} $S(k_{\text{Bragg}})$ is close to the pristine case. However, under the FC protocol, the system settles itself in a configuration such that $S(k_{\text{Bragg}})$ is practically zero. {In addition, Fig. \subref{fig1d} shows that, for the Abrikosov initial configuration, the behavior of the average of $\Psi_6$ is quite similar to that of $S(k_{\text{Bragg}})$ in Fig. \subref{fig1c}.} {Remarkably}, Fig. \subref{fig1d} shows that, under the FC protocol{, and low temperatures,} the average {of the magnitude of $\Psi_6$ stabilizes at two values: a negative value with $|\Psi_6|\approx0.6$ (26 samples); and a positive value with $|\Psi_6|\approx0.2$ (4 samples; for these samples, $\Psi_6$ can take positive or negative values as $T$ increases up to the melting transition). These results suggest} that the local order is preserved, and the BrG phase sets in for this value of $B_\phi$.

Lastly, Figs. \subref{fig1e} and \subref{fig1f} present $S(k_{\text{Bragg}})$ and $\Psi_6$, respectively, {at} $B_\phi=B/32$, a very low concentration of defects. Figure \subref{fig1e} shows that, for the Abrikosov initial configuration, the low concentration of defects does not affect much the $T$-dependence of $S(k_{\text{Bragg}})$ relative to {the pristine curve}. {Likewise, Fig. \subref{fig1f} also shows that there is no significant difference between $\Psi_6$ using the Abrikosov initial configuration and a pristine sample. Most importantly, we stress that the data of $S(k_{\text{Bragg}})$ under the FC protocol exhibit two average values at low temperatures: $S(k_{\text{Bragg}})\approx0$ (16 samples) and $S(k_{\text{Bragg}})\approx0.95$ (13 samples); one sample was excluded on physical grounds. Further, the data for the average FC values of $\Psi_6$ in Fig. \subref{fig1f} also exhibit a distinct behavior for the samples mentioned above: for the 16 samples with $S(k_{\text{Bragg}})\approx0$, we find that 11 samples exhibit low-temperature values of $\Psi_6$ close to $-0.8$, while 5 samples display values close to $-0.3$. On the other hand, for those samples with $S(k_{\text{Bragg}})\approx0.95$,  the average value of $\Psi_6$ is close to one. The system is in a BrG phase with very intriguing features.}

\begin{figure}[htp]
\begin{center}
\subfloat{\includegraphics[width=0.23\textwidth]{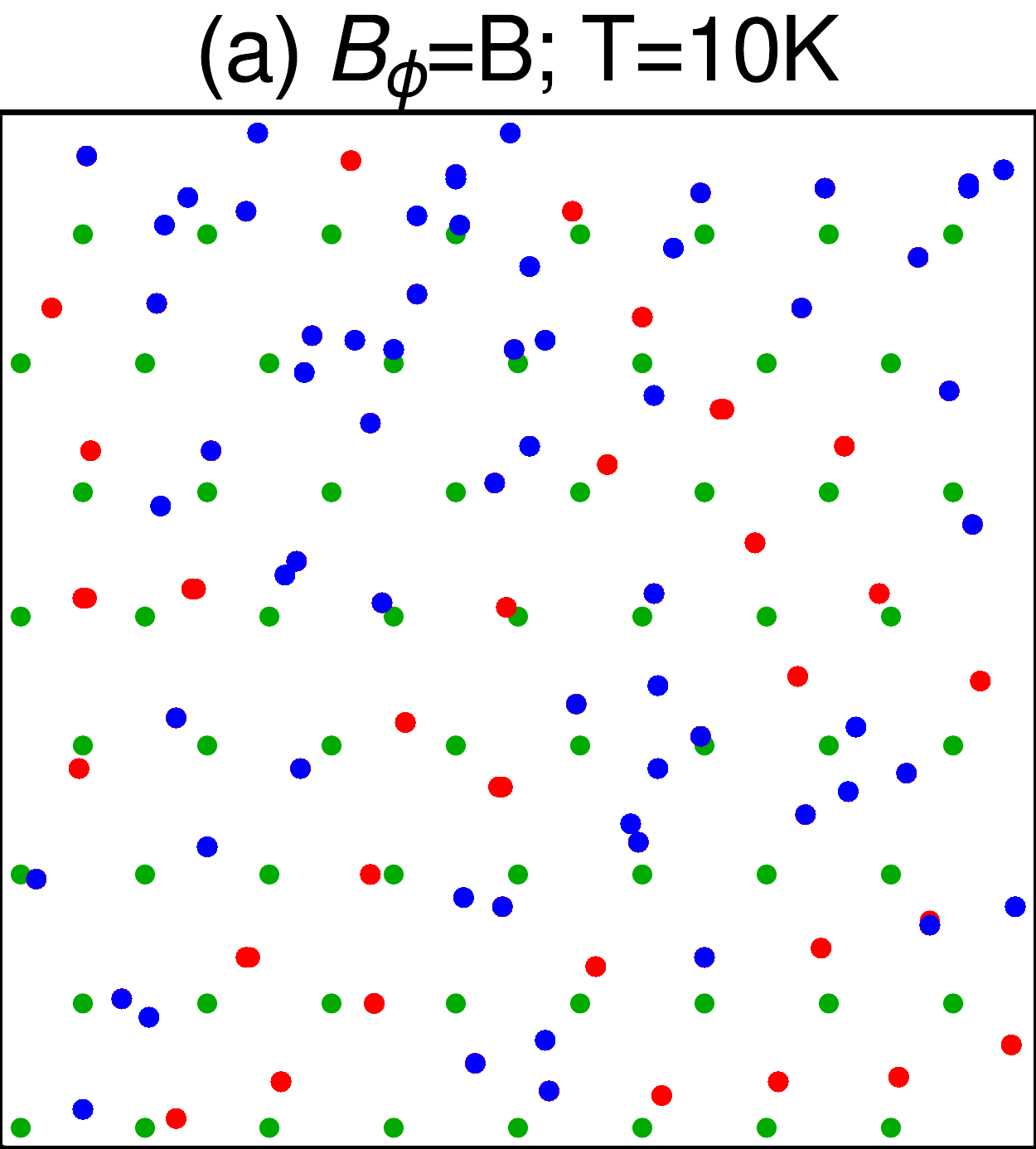}\label{fig2a}}
\subfloat{\includegraphics[width=0.23\textwidth]{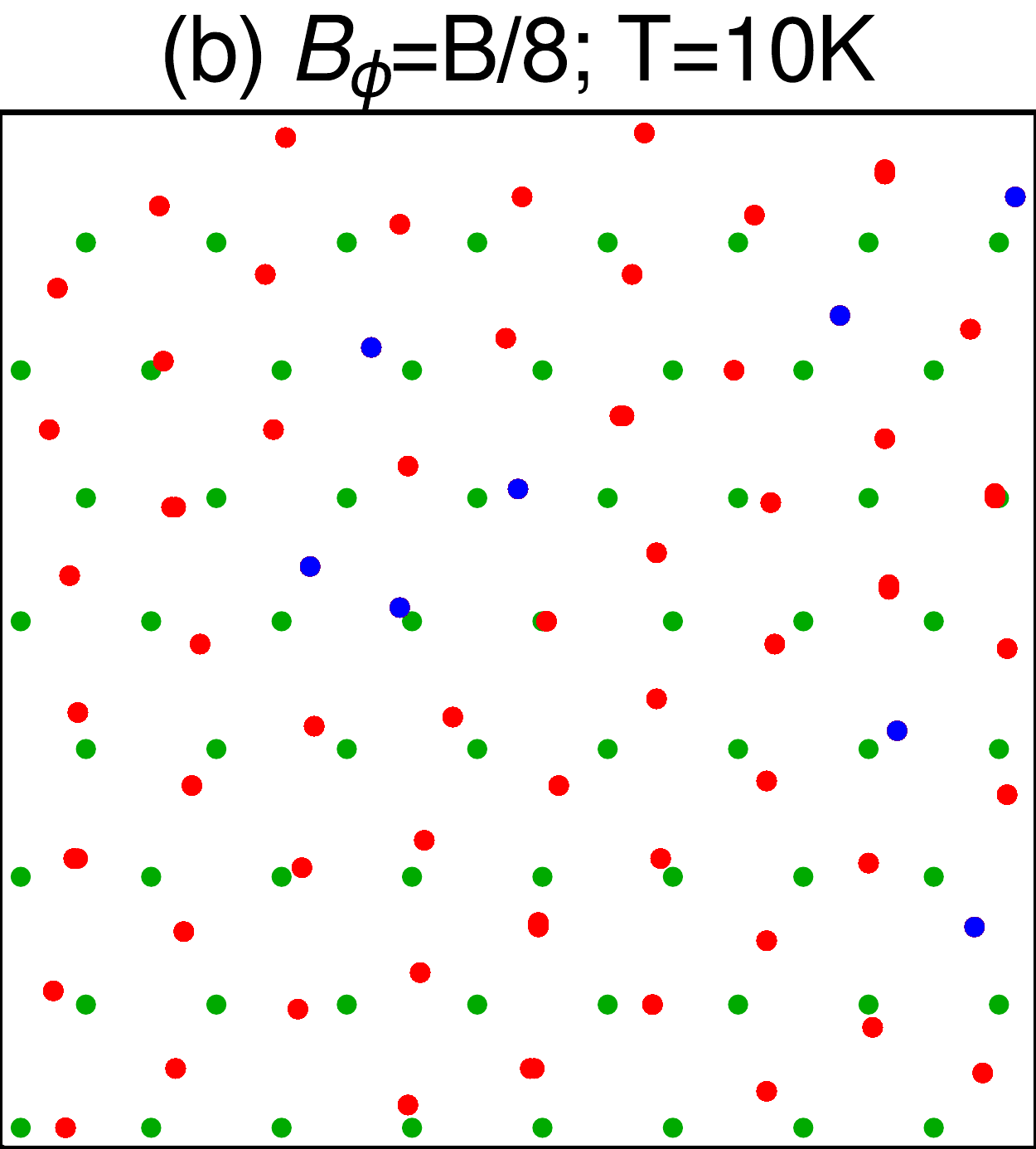}\label{fig2b}}\\
\subfloat{\includegraphics[width=0.23\textwidth]{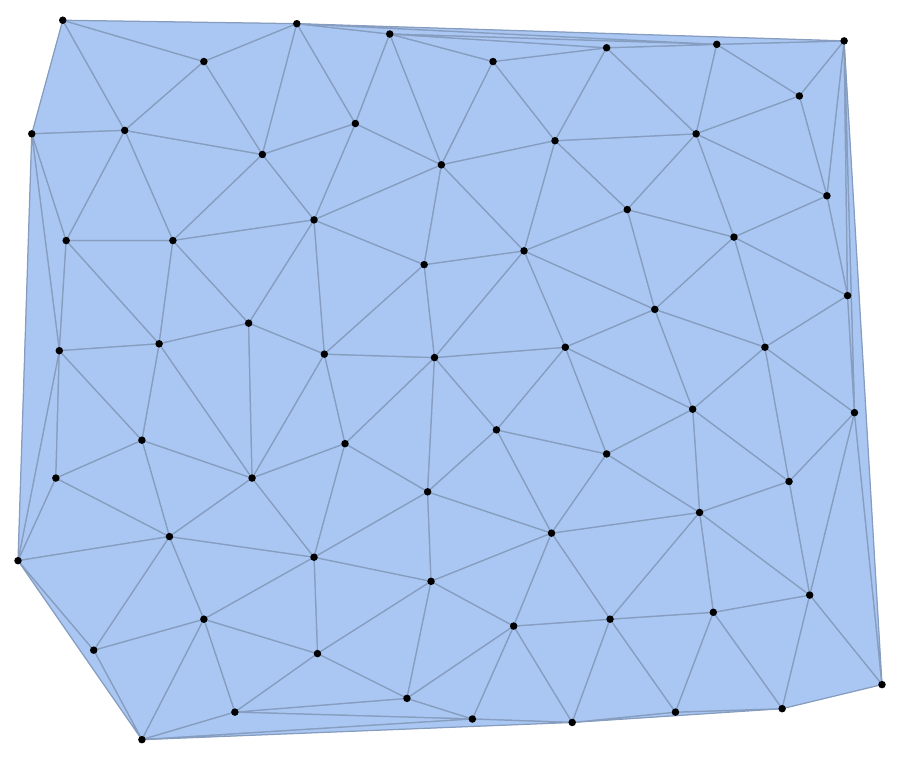}\label{fig2c}}
\subfloat{\includegraphics[width=0.23\textwidth]{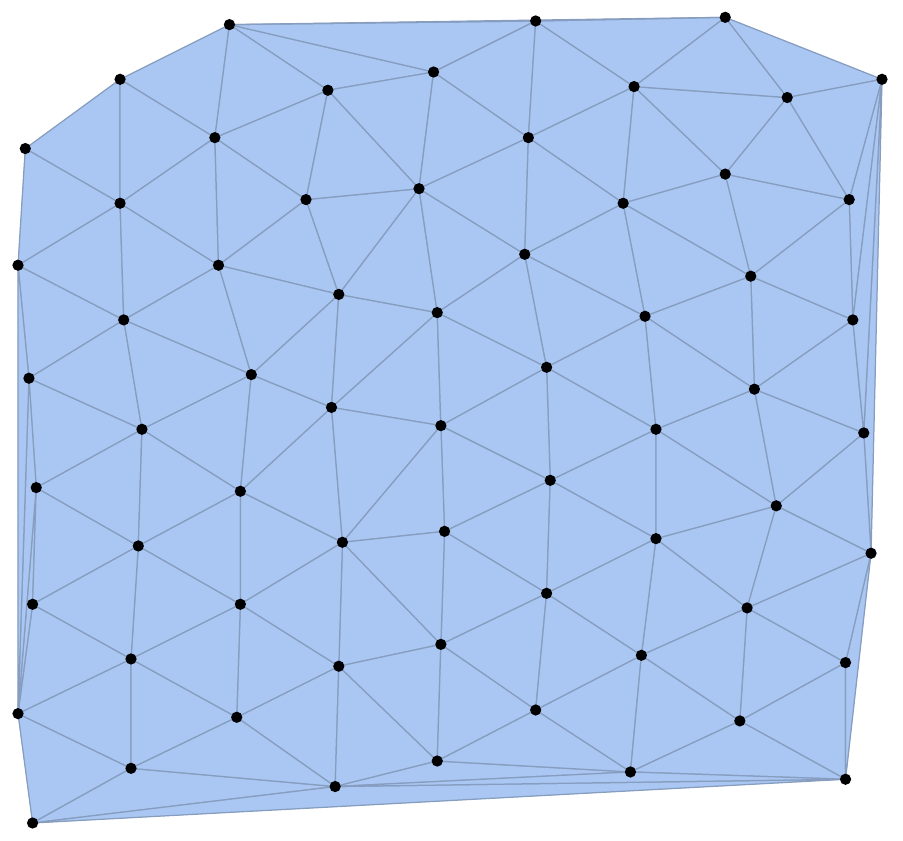}\label{fig2d}}\\
\subfloat{\includegraphics[width=0.23\textwidth]{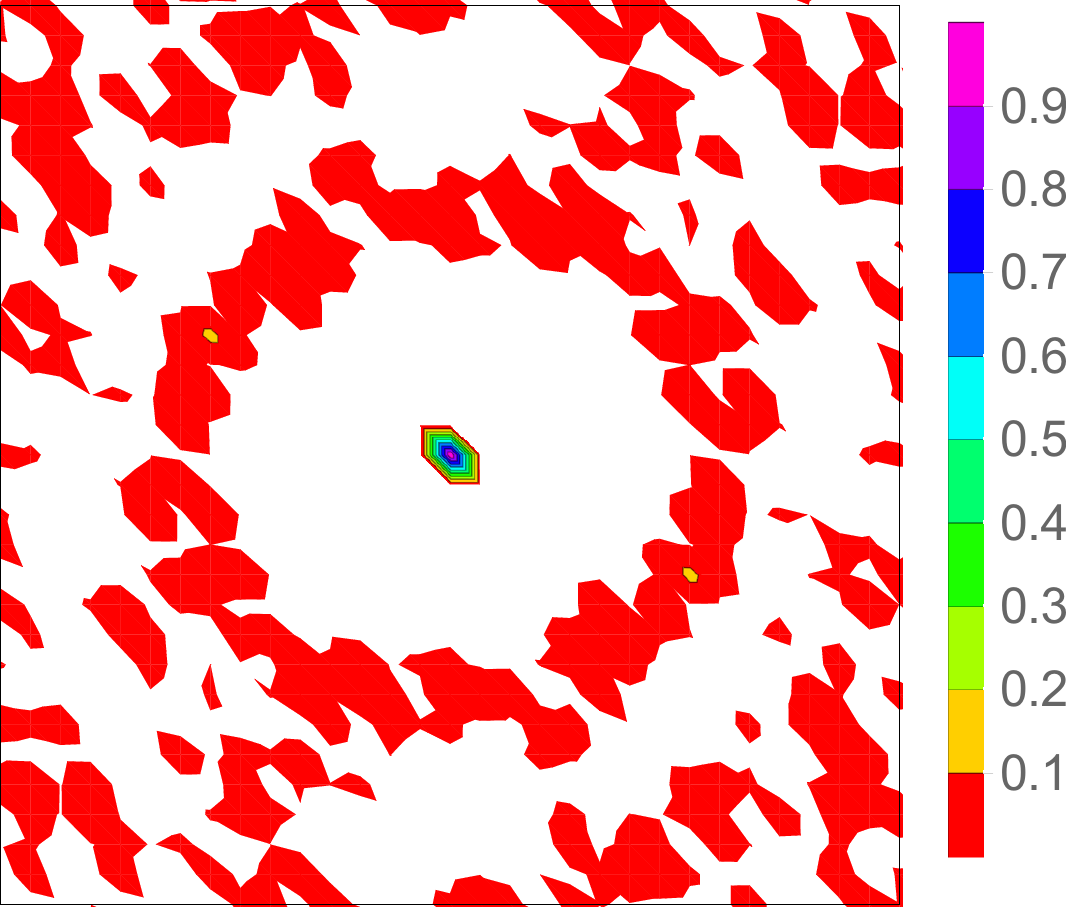}\label{fig2e}}
\subfloat{\includegraphics[width=0.23\textwidth]{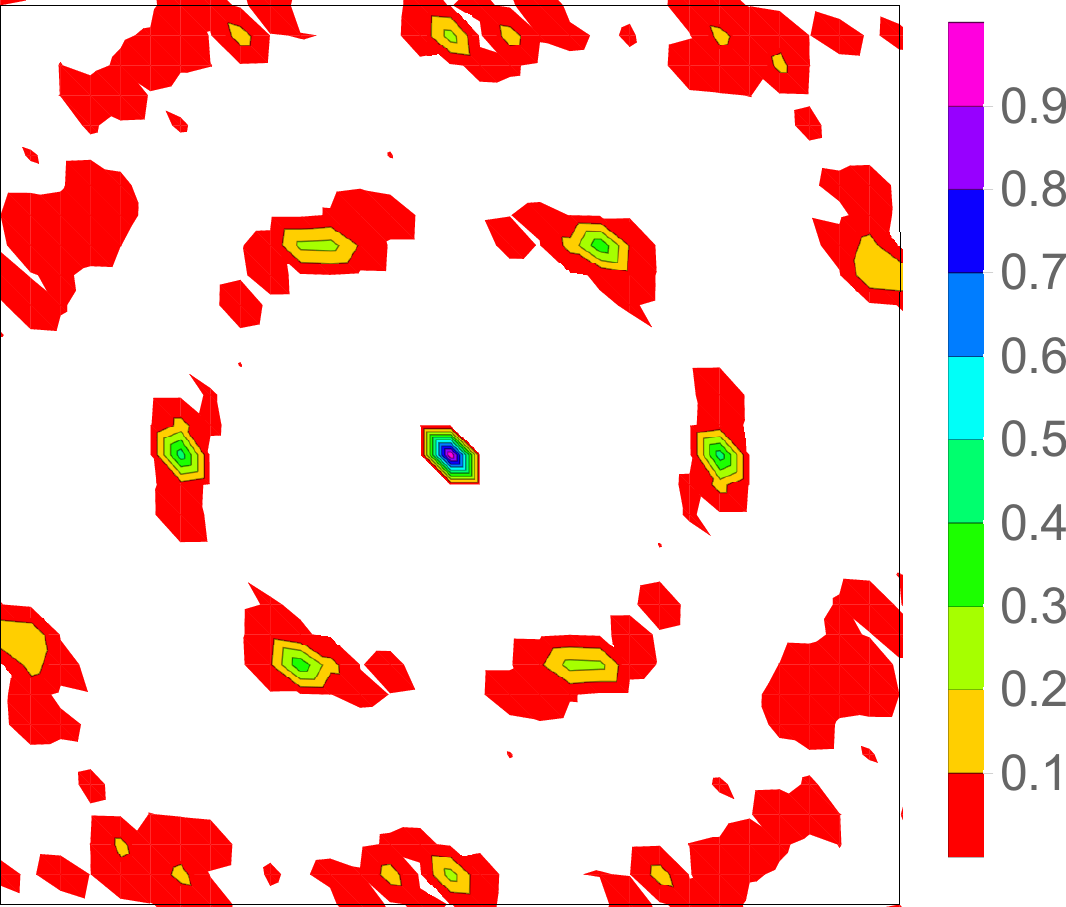}\label{fig2f}}\\
\subfloat{\includegraphics[width=0.23\textwidth]{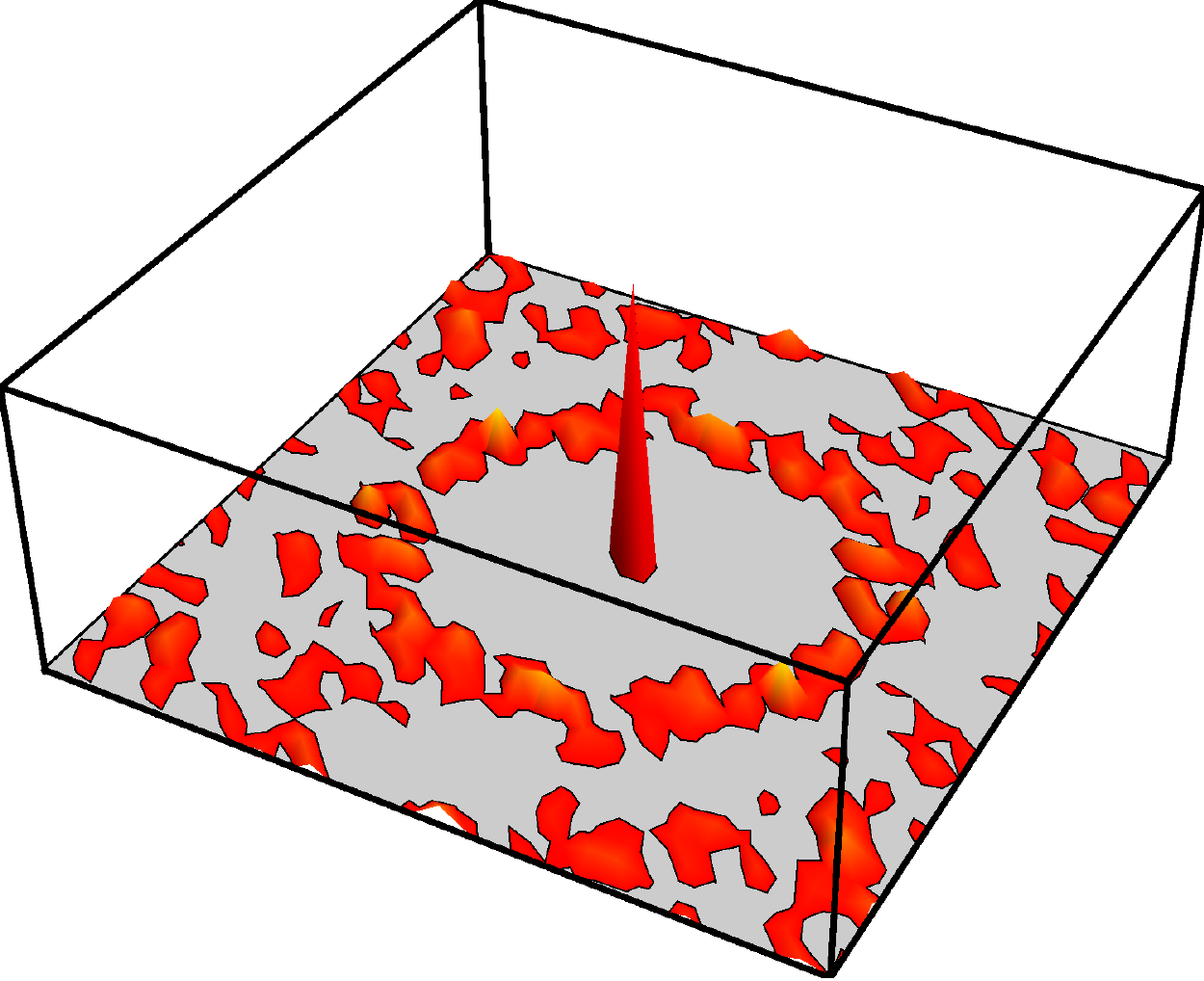}\label{fig2g}}
\subfloat{\includegraphics[width=0.23\textwidth]{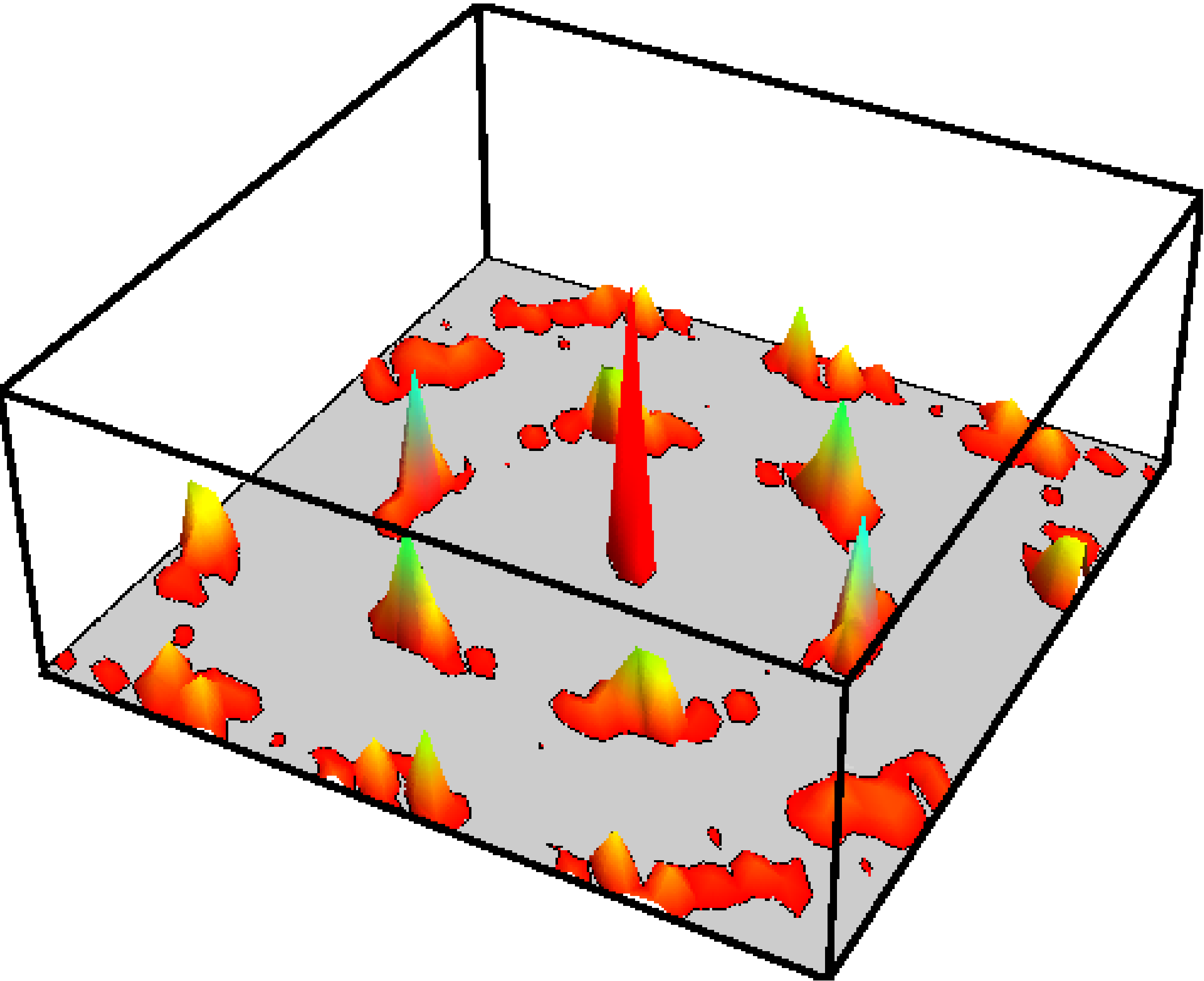}\label{fig2h}}\\
\caption{(Color online) {Snapshots of the} planar projection of  typical vortex lattice configurations, columnar defects and {ideal} Abrikosov lattice, indicated in the first row by red, blue, and green dots, respectively at $B=0.1$ kG and $T=10$ K for FC samples starting at $T_0=80$ K. The second row displays Delaunay triangulation of the planar projection of the vortex lattice. {The third and fourth rows} exhibit the planar structure factor {$S(k_\bot, z=N_z/2)$} of the vortex lattice {and the corresponding Bragg peaks, respectively}. Data were obtained from samples, with (a) $B_\phi=B$ (first column) and (b) $B_\phi=B/8$ (second column).\label{fig2}}
\end{center}
\end{figure}

In order to unveil the effect of the presence of short or long range order on FC samples {as the system crossovers form th BG to the BrG phase}, we must observe the actual configuration of vortices. In this context, the first row of Fig. \ref{fig2} exhibits {snapshots of typical vortex configurations with the average of $S(k_\text{Bragg})$ nearly zero at $B=0.1$ kG} and $T=10$ K{, a temperature much lower than the melting temperature,} for $B_\phi=B$ [Fig. \subref{fig2a}] and $B_\phi=B/8$ [Fig. \subref{fig2b}], respectively. 

In Fig. \subref{fig2a}, we observe that the lattice is indeed disordered, with many vortices (in red) pinned to the defects (in blue), making the configuration diverge from the ideal Abrikosov lattice (in green). {Under this condition, we stress that many red pinned vortices do not appear explicitly because the blue color of the deffects dominates when their planar projections coincide.} This can be corroborated by the Delaunay triangulation in the first diagram of the second row, which suggests the occurrence of many vortices with coordination number equal to five or seven, instead of six. {Lastly,} in the third {and fourth rows}, Fig. \subref{fig2a} shows that the system does not have sharp Bragg peaks at this temperature, but rather displays the characteristics of a BG phase.

Figure \subref{fig2b} shows that, {at} $B_\phi=B/8$, though some vortices are pinned, the system resembles a \textit{reoriented} hexagonal lattice.  In the second diagram of the second row, we observe some distortions relative to the ideal hexagonal lattice. There are some vortices with coordination number different from six, although they are relatively isolated within the lattice. In the {third and fourth rows,} the structure factor {and the power law decay of the Bragg peaks are displayed, respectively. These results evidence} that we have reached, within numerical precision, the onset of the BrG phase.

\begin{figure}[htp]
\begin{center}
\subfloat{\includegraphics[width=0.23\textwidth]{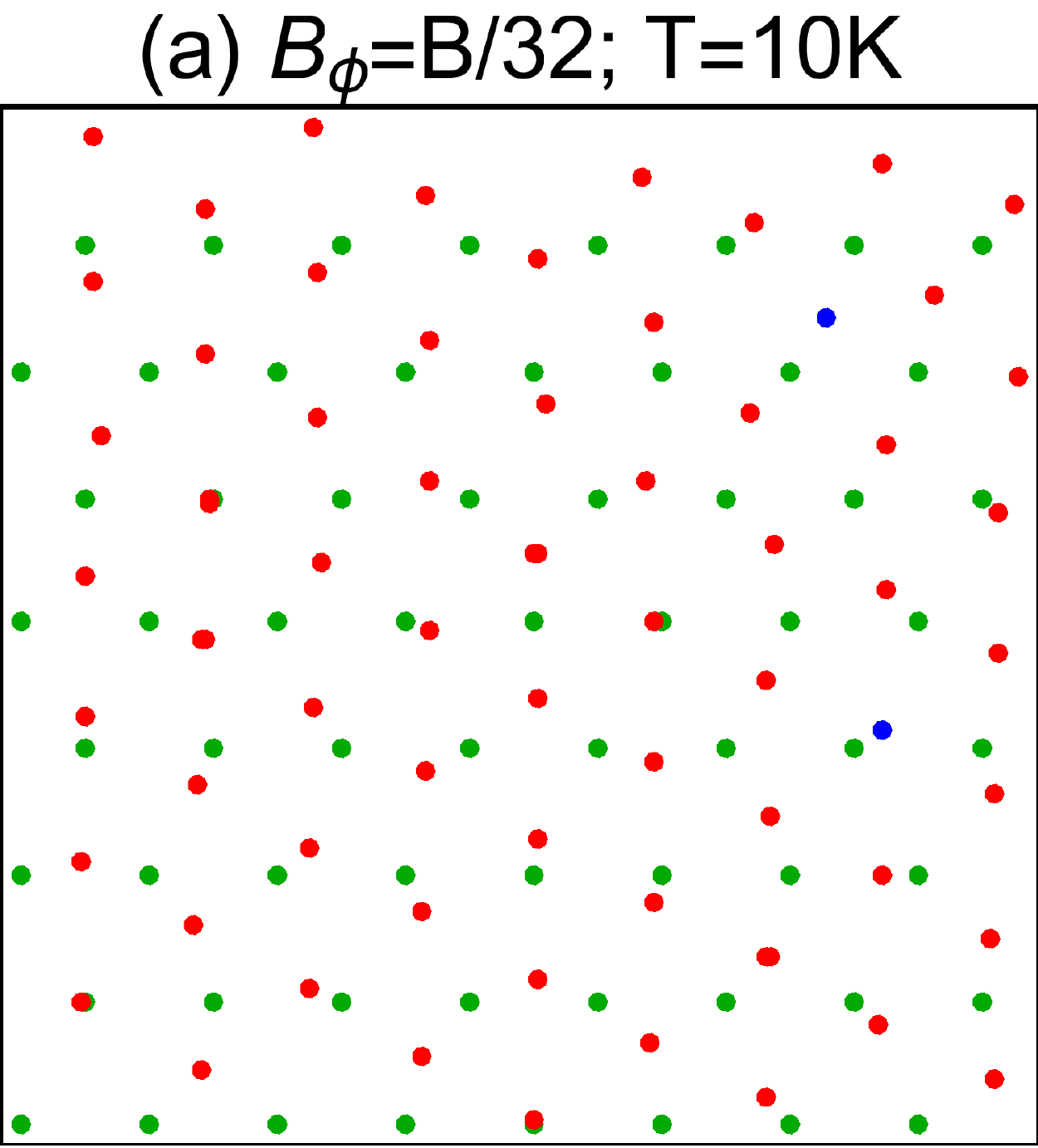}\label{fig3a}}
\subfloat{\includegraphics[width=0.23\textwidth]{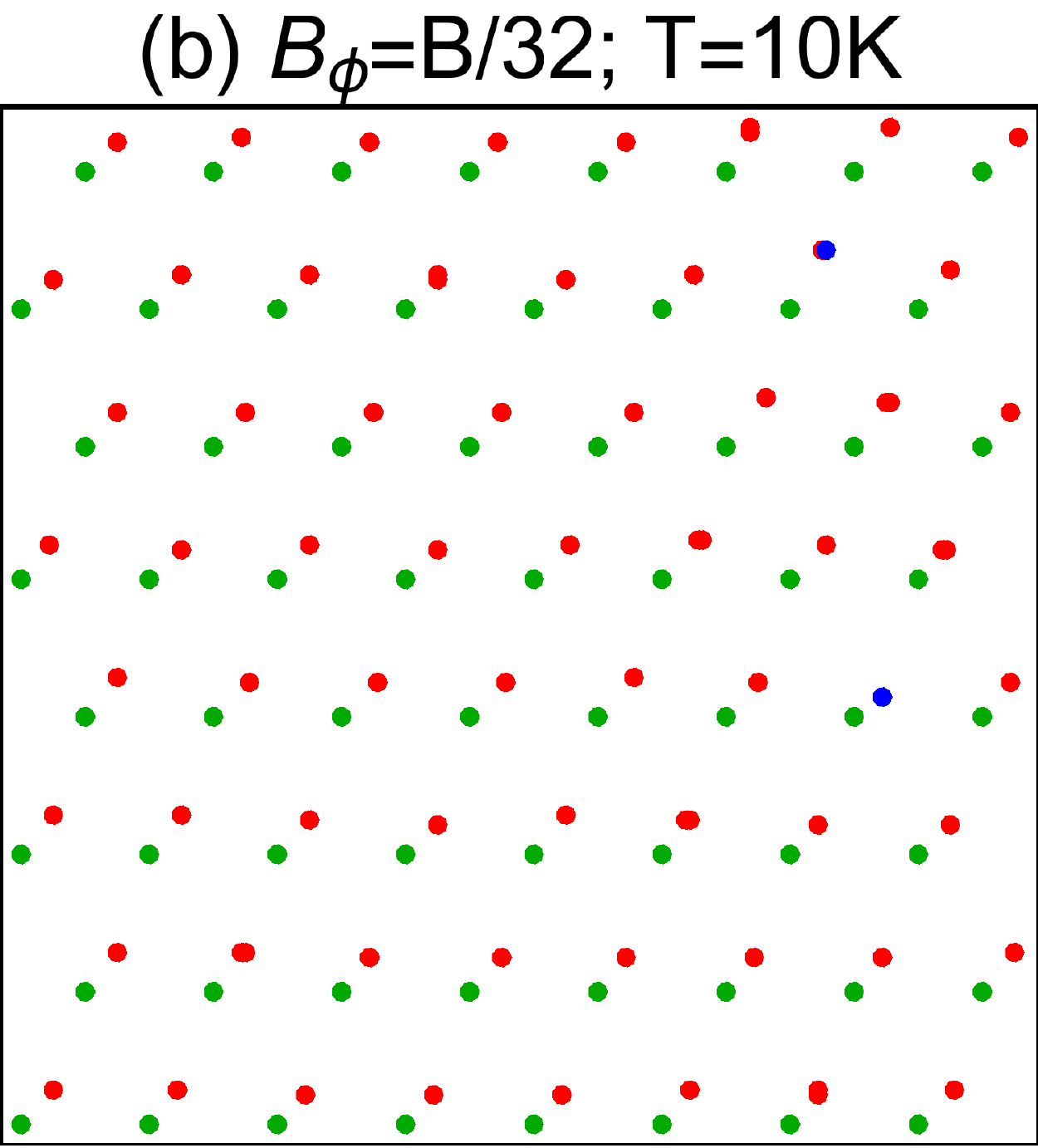}\label{fig3b}}\\
\subfloat{\includegraphics[width=0.23\textwidth]{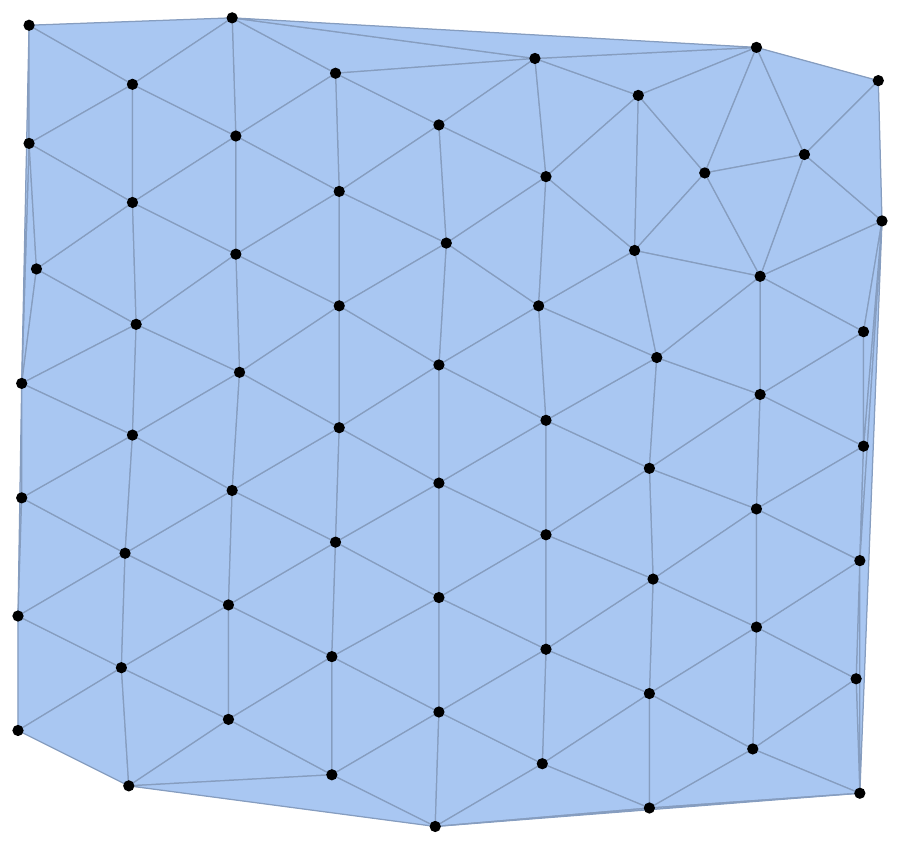}\label{fig3c}}
\subfloat{\includegraphics[width=0.23\textwidth]{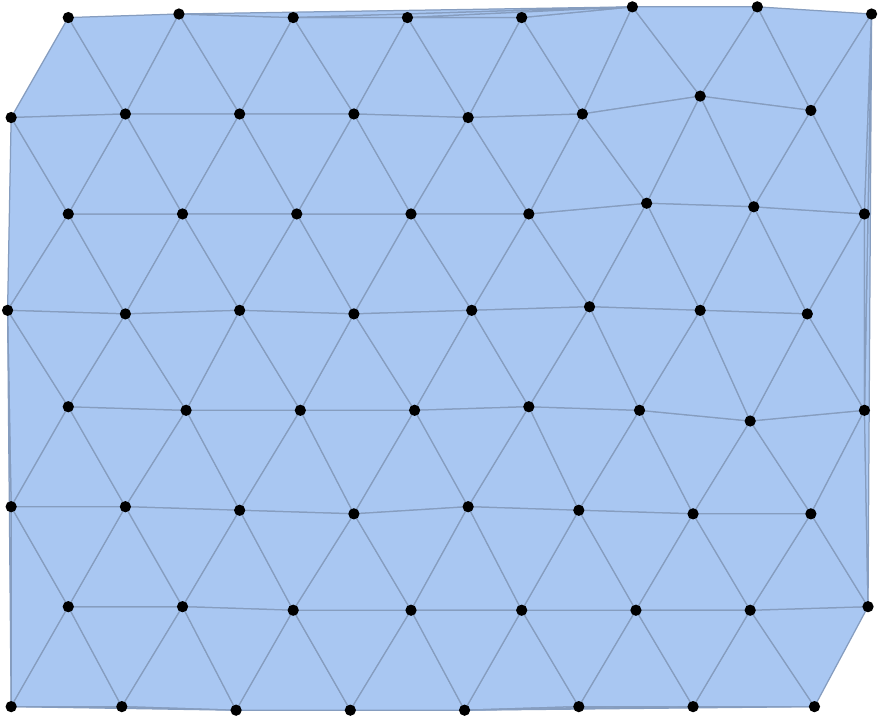}\label{fig3d}}\\
\subfloat{\includegraphics[width=0.23\textwidth]{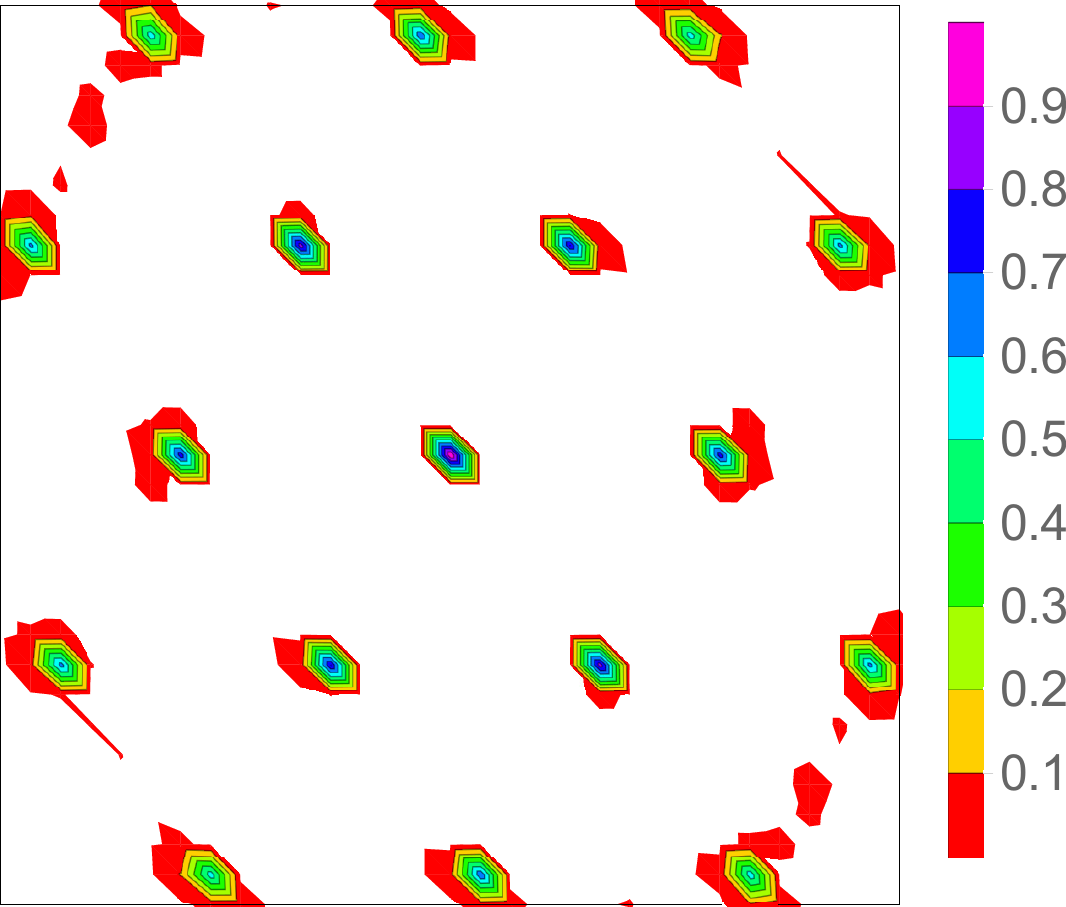}\label{fig3e}}
\subfloat{\includegraphics[width=0.23\textwidth]{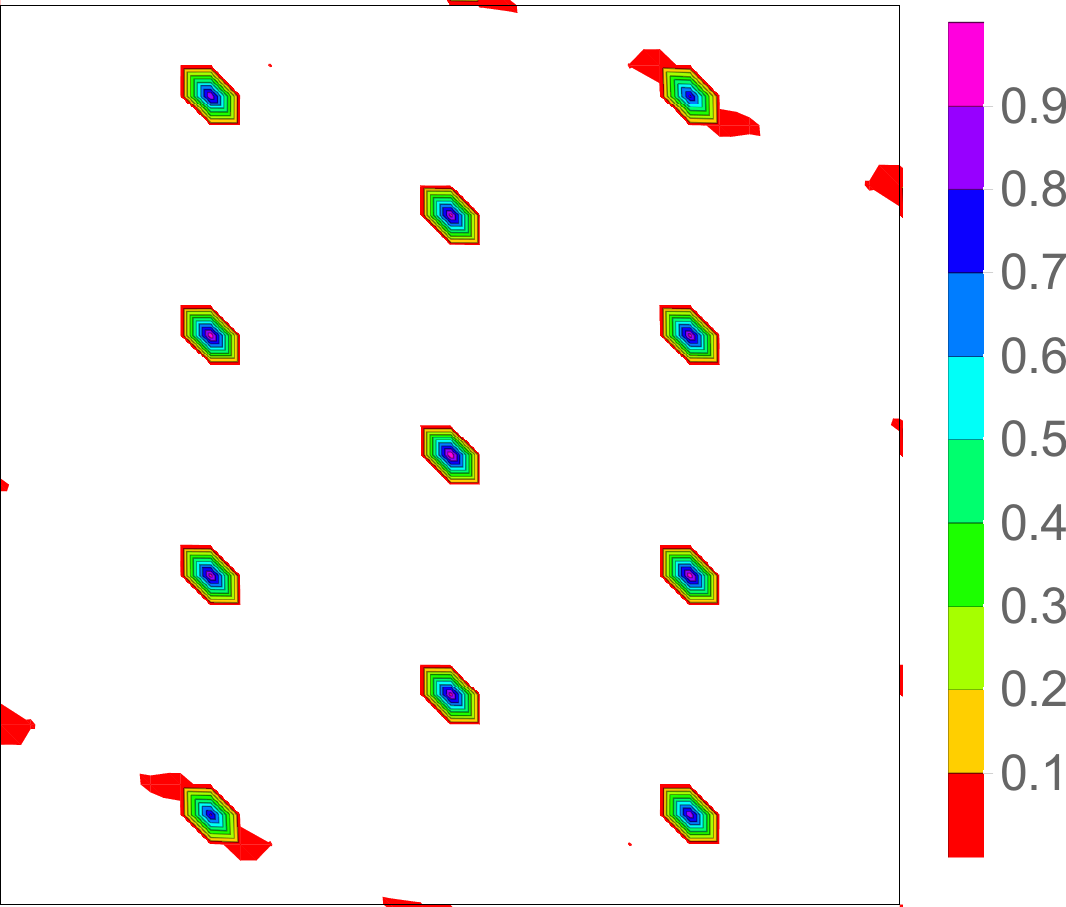}\label{fig3f}}\\
\subfloat{\includegraphics[width=0.23\textwidth]{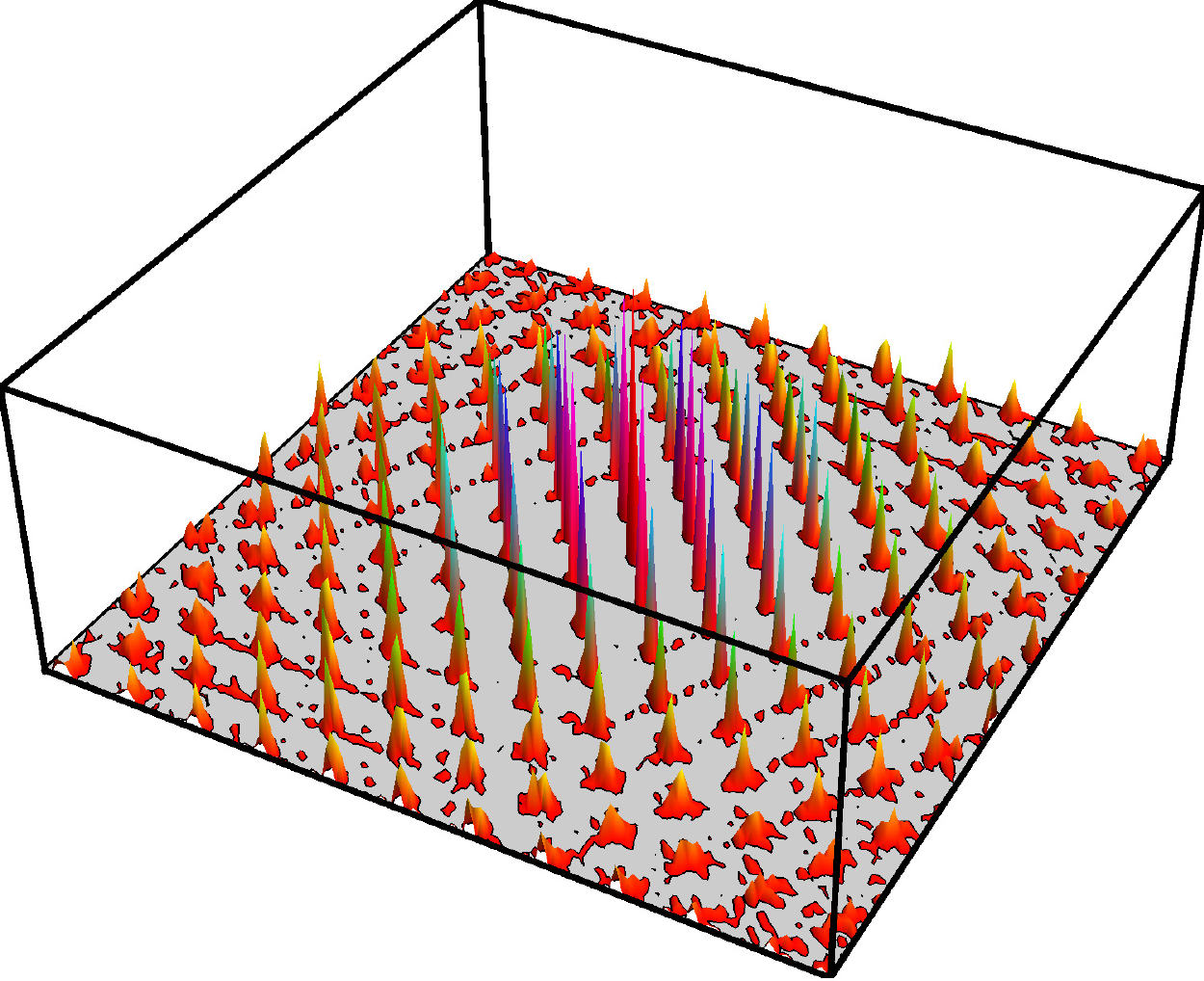}\label{fig3g}}
\subfloat{\includegraphics[width=0.23\textwidth]{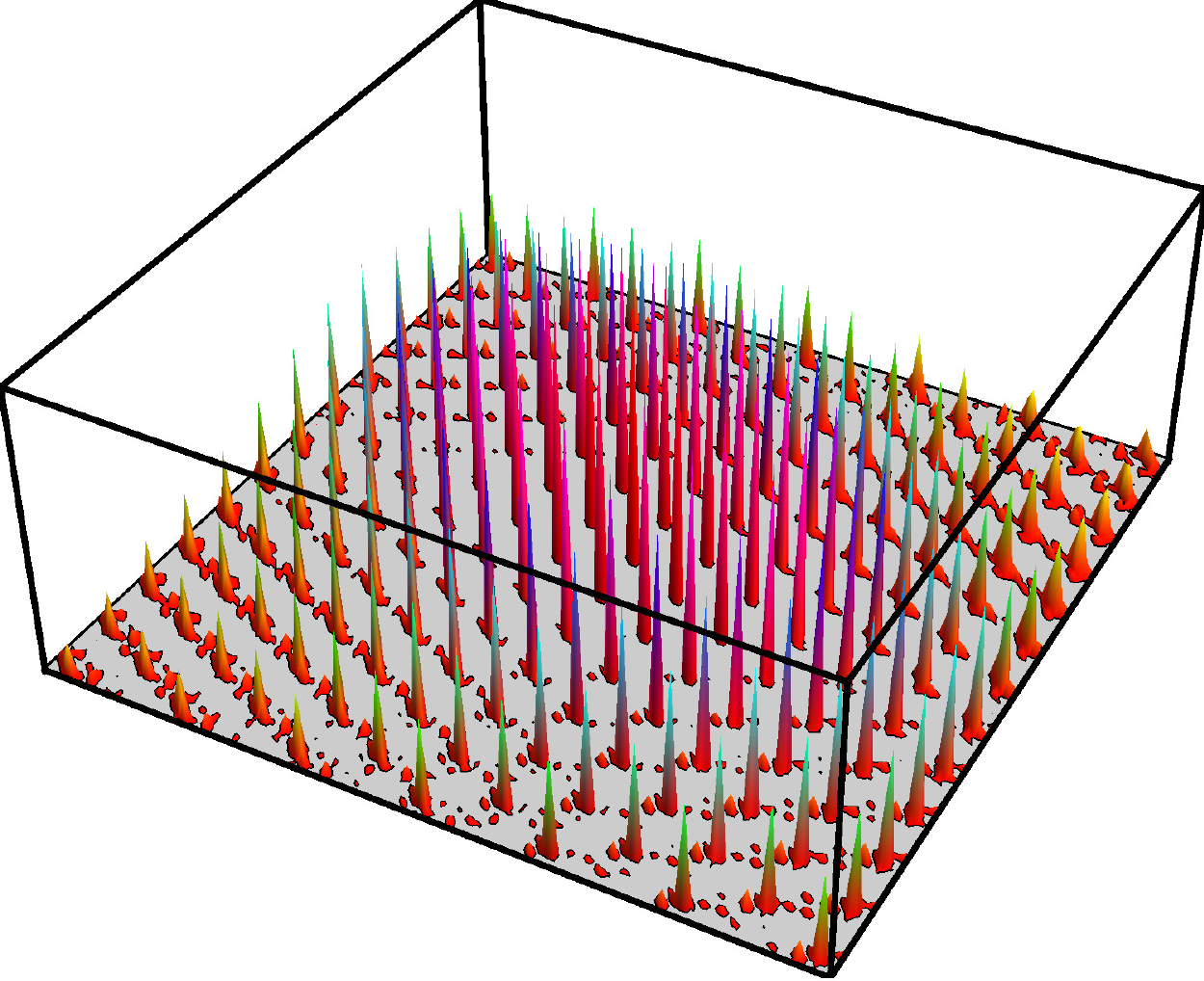}\label{fig3h}}\\
\caption{{(Color online) Snapshots of the planar projection of  typical vortex lattice configurations, columnar defects and ideal Abrikosov lattice, indicated in the first row by red, blue, and green dots, respectively at $B=0.1$ kG and $T=10$ K for FC samples starting at $T_0=80$ K. The second row displays Delaunay triangulation of the planar projection of the vortex lattice. The third and fourth rows exhibit the planar structure factor $S(k_\bot, z=N_z/2)$ of the vortex lattice and the corresponding Bragg peaks, respectively. Data were obtained from samples, with: (a) $B_\phi=B/32$ and $S(k_\text{Bragg})\approx0$ (first column); (b) $B_\phi=B/32$ and $S(k_\text{Bragg})\approx0.95$ (second column).}\label{fig3}}
\end{center}
\end{figure}

{We stress that negative values of $\Psi_6$ are due to the rotation of the lattice with respect to the ideal Abrikosov lattice, as shown in in the first row of Fig. \subref{fig3a}, at $B_\phi=B/32$. On the other hand, for $\Psi_6\approx1$, the lattices are translated with respect to the ideal Abrikosov lattice, as shown in the first row of Fig. \ref{fig3}, also at $B_\phi=B/32$. Delaunay triangulations are also shown in the second row of Figs. \subref{fig3a} and \subref{fig3b}. Lastly, in the third and fourth rows of Fig. \ref{fig3}, we show the planar projections and the corresponding Bragg peaks, which confirm the global effect of the presence of disorder: rotation [Fig. \subref{fig3a}] or displacement [Fig. \subref{fig3b}] with respect to the ideal Abrikosov lattice, thus suggesting the power law decay behavior of the Bragg peaks in both cases.}

\section{BrG phase and divergent Bragg peaks\label{sec:4}}

The results of the previous section gave us evidence of the presence of a BrG phase in our simulations. Nonetheless, to confirm it, we must verify that the structure factor diverges algebraically as the wave vector $k\rightarrow0$, or, equivalently, that the correlation function decreases more slowly than a standard exponential decay at large distances. However, since our system has a finite size, it imposes us a limitation: the wave vector resolution is limited to $\Delta k=2\pi/L$, where $L$ is the size of the system. Here, we use an approach compatible with that of Giamarchi and Le Doussal \cite{giamarchi:1994, *giamarchi:1995}, and also employed by Klein \textit{et al.} \cite{klein:2001} to identify the BrG phase from neutron diffraction data from a \textit{isotropic} single-phase (K, Ba)BiO${}_3$ crystal at $T=2$ K. In fact, the experimentally observed lorentzian behavior of the structure factor observed for various field values implies a decrease of the central peak with increasing field, while the width of the curve does not change. Instead, in our numerical-analytical description, the field and temperature are fixed, while the density of columnar defects decreases, and the numerical data is used as input. The width of the lorentzian curve is in fact limited by the experimental resolution or by the numerical accuracy.

Let ${\bm r}(z)=(x(z), y(z))$ be the position of a given vortex in the $z$ layer. We define the root mean square deviation of the position of the vortices along the $z$-direction as

\begin{equation}\label{eq:11}
 \Delta r^2_{rms}(\lvert z-z'\rvert)=\overline{\langle \lvert{\bm r}(z)-{\bm r}(z')\rvert^2\rangle},
\end{equation}
where the angular bracket represents the Boltzmann thermal average and the upper bar the disorder average. On the other hand, we define the 3D {density-density vortex} correlation function $G$  in terms of its Fourier transform, i. e., the 3D structure factor $S(\bm k)$:
\begin{equation}\label{eq:12}
 G(\lvert \bm r-\bm r'\rvert)=\frac{V}{N}\int S(\bm k)e^{-i\bm k_{\bot}\cdot(\bm r-\bm r')}e^{-ik_z(z-z')}\text{d}^3k,
\end{equation}
where $V$ is the volume of the system, and $N$ is the total number of vortices. We can estimate the spatial dependence of $G$ by taking the saddle point of Eq. \eqref{eq:12} around the Bragg peaks and under disorder average. We thus retain only the Bragg peaks average:
\begin{equation}\label{eq:13}
 G(\lvert \bm r-\bm r'\rvert)\simeq \frac{8\pi^3}{N}\sum_{\bm K} S(\bm K)\overline{\langle e^{-i\bm K_\bot\cdot(\bm r-\bm r')}\rangle}.
\end{equation}
Note that \eqref{eq:13} has no explicit dependence on $z$; indeed, since the $z$-component of the Bragg vector $K_z=2\pi m/d$ and $z-z'=nd$, where $n$ and $m$ are integers, $e^{iK_z(z-z')}=1$. In addition, the right hand side of \eqref{eq:13} must be real, thereby
\begin{equation}\label{eq:14}
 G(\lvert \bm r-\bm r'\rvert)\simeq \frac{8\pi^3}{N}\sum_{\bm K} 2S(\bm K)\overline{\langle \cos\left(\bm K_\bot\cdot(\bm r-\bm r')\right)\rangle};
\end{equation}
and at low temperatures {(much lower than the melting temperature)}, we can safely write:
\begin{equation}\label{eq:15}
 G(\lvert \bm r-\bm r'\rvert)\simeq \frac{8\pi^3}{N}\sum_{\bm K} 2S(\bm K)\overline{\left\langle 1-\frac{K_\bot^2}{2}\lvert\bm r(z)-\bm r(z')\rvert^2\right\rangle},
\end{equation}
which implies
\begin{equation}\label{eq:16}
 G(\lvert \bm r- \bm r'\rvert)\simeq C_1-C_2\Delta r^2_{rms}(\lvert z-z'\rvert),
\end{equation}
where $C_1$ and $C_2$ are {$z$-independent functions}. 

The results above show that we can use the root mean square deviation of the position of the vortices in order to estimate the decay of their correlation function along the $z$ direction. We thus retrieve the dependence of the structure factor on $k_z$, through the following Fourier transform:
\begin{equation}\label{eq:17}
 \sigma(k_z)=\left\lvert\frac{1}{d}\int_0^{N_zd} \Delta r^2_{rms}(z-z')e^{ik_z(z-z')}\text{d}(z-z')\right\rvert,
\end{equation}
{where $N_zd$ is the sample size, $k_zd=2\pi m/N_z$, $m$ is an integer, with $\Delta r^2_{rms}(z-z')$ satisfying the boundary condition: $\Delta r^2_{rms}(0)=\Delta r^2_{rms}(N_zd)$.} As the planes are discretized, with a distance $d$ between each other, the integral of \eqref{eq:17} can be transformed into a sum over all planes of the sample:
\begin{equation}\label{eq:18}
 \sigma(k_z)=\left\lvert\sum_{n=0}^{N_z-1} \Delta r^2_{rms}e^{ink_zd}\right\rvert.
\end{equation}

\begin{figure}[htp]
\begin{center}
\hspace{-5mm}\includegraphics[width=0.5\textwidth]{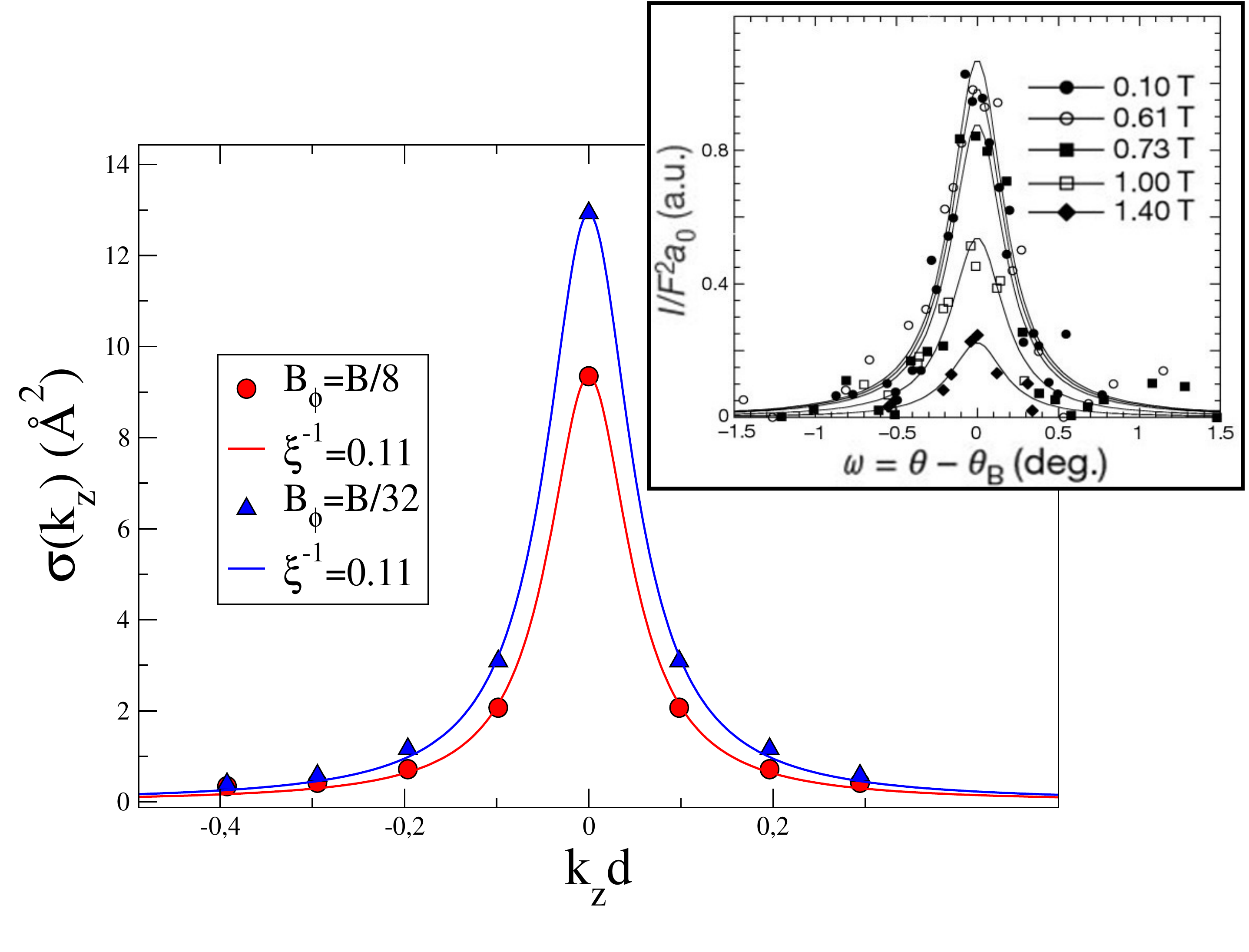}
\caption{(Color online) $\sigma(k_z)$, defined by Eq. \ref{eq:18}, {for two typical FC samples} at {$B=0.1$ kG and $T=10$ K} for $B_\phi=B/8$ (red dots) and $B_\phi=B/32$ (blue triangles). The curves represent lorentzian fits for each data set. Both fits have width $d/\xi\simeq 0.11$, very close to the numerical resolution $\Delta k_zd=2\pi/64\simeq0.1$. For comparison, the inset shows experimental data of the neutron beam intensity diffracted by vortices in a single-phase (K, Ba)BiO${}_3$ crystal taken from Ref. \onlinecite{klein:2001}.  \label{fig4}}
\end{center}
\end{figure}

We have calculated $\Delta r^2_{rms}$ for samples at $T=10$ K, averaged over columnar disorder, and $B_\phi=B/8$ and $B_\phi=B/32${, using MC data from simulations performed in Section III}. We have also calculated the discrete Fourier transform $\sigma(k_z)$ of $\Delta r^2_{rms}$ and fitted the data to a lorentzian, in the context of the Ornstein-Zernike framework \cite{stanley:1987}:
\begin{equation}\label{eq:19}
 \sigma(k_z)=\frac{A}{(\xi^{-1}/2)^2+k_z^2},
\end{equation}
where {$A$ is an adimensional $k_z$-independent function and} $\xi^{-1}$ is the width of the lorentzian. As $N_z\rightarrow\infty$, $\xi^{-1}\rightarrow0$ and $\sigma(k_z)\sim k_z^{-2}$, which characterizes an algebraic divergence at $k_z=0$. However, our minimum fitting value of $d/\xi$ is limited by the numerical precision $\Delta k_zd=2\pi/N_z=2\pi/64\approx 0.1$. Figure \ref{fig4} displays the results of $\sigma(k_z)$ for the two distinct averaged samples.  Although the peaks have different heights, the width of the lorentzian fit is the same, a signature of a BrG phase. In fact, the fitting value for the two samples is $d/\xi\simeq 0.11$, which is very close to the numerical resolution. {In order to test the dependence of $d/\xi$ on $N_z$, we have calculated it for a sample with $N_z=256$, which corresponds to $\Delta k_zd=2\pi/256\approx0.025$ and a width $d/\xi\approx0.03$, in very good agreement with the power-law decay of the divergent Bragg peaks in the BrG phase, as predicted by Eq. \eqref{eq:19}}. We believe this is a conclusive evidence of the emergence of the BrG phase in our simulations at low fields and low columnar disorder with BSCCO parameters. For comparison, the inset in Fig. \ref{fig4} shows the experimental data \cite{klein:2001} of the neutron beam intensity $I$ diffracted by vortices in the isotropic single-phase (K, Ba)BiO${}_3$ crystal, where $F$ is the single-vortex standard form factor, $a_0$ is the lattice spacing, and $\omega=\theta-\theta_B$ measures the mismatch of the angle $\theta$, where $\theta$ is the angle between the magnetic field and the neutron beam, and $\theta_B$ is the Bragg angle.

\section{Concluding Remarks\label{sec:5}}

In this work, we presented simulations of \textit{highly anisotropic} layered samples with BSCCO parameters, in the low field regime with focus on the observation of the crossover from BG to BrG phase as the density of columnar defects is lowered.

Under the above conditions, the vortices tend to stay in a glassy phase with sharp Bragg peaks. In particular, using the field-cooling {(FC)} protocol, we verified that the {magnitude of the} hexatic order parameter $\Psi_6$ is {greater or of the same order of} $S(k_\text{Bragg})$, which implies the prevalence of short range over long range order. 

Notably, by examining the behavior of the 3D structure factor along the $z$ direction by means of a relation between the {density-density vortex} correlation function and the root mean square deviation of the positions of the vortices, we found that it fits to a lorentzian function close to $k_z=0$, with the same width for the simulated samples {in the BrG phase}. This feature is a signature of the occurrence of a BrG phase under low fields and low densities of columnar defects, and provides a clear-cut demonstration of the algebraic divergence of the Bragg peaks as the wave vector along the field direction $k_z\rightarrow0$, in the context of Ornstein-Zernike framework. {In conclusion, our reported numerical and analytical results have unveiled several intriguing features of vortex matter as it crossovers from BG to BrG phase at low fields and low concentration of disorder.}

\section*{acknowledgments}

This work was supported by CNPq and FACEPE through the PRONEX program, and CAPES (Brazilian agencies).

\end{document}